\pdfoutput=1
\documentclass{mn2e}
\usepackage[dvipsnames,usenames]{color}
\usepackage{colortbl}
\usepackage{epsf,times}
\usepackage{graphicx}
\usepackage{amsmath}
\let\bibhang\relax

\usepackage{natbib}

\begin{document}
\title[Numerical modelling of radio galaxy lobes]{Numerical modelling
  of the lobes of radio galaxies in cluster environments}
\author[M.J.\ Hardcastle \& M.G.H.\ Krause]
{M.J.\ Hardcastle$^1$ and M.G.H.\ Krause$^{2,3}$\\
$^1$ School of Physics, Astronomy and Mathematics, University of
  Hertfordshire, College Lane, Hatfield AL10 9AB\\
$^2$ Excellence Cluster Universe, Technische Universit\"at M\"unchen,
  Boltzmannstrasse 2, 85748 Garching, Germany\\
$^3$ Max-Planck-Institut f\"ur extraterrestrische Physik, Postfach
  1312, Giessenbachstrasse, 85741 Garching, Germany\\
}
\maketitle
\begin{abstract}
We have carried out two-dimensional, axisymmetric, hydrodynamic
numerical modelling of the evolution of radio galaxy lobes. The
emphasis of our work is on including realistic hot-gas environments in
the simulations and on establishing what properties of the resulting
radio lobes are independent of the choice of environmental properties
and of other features of the models such as the initial jet Mach
number. The simulated jet power we use is chosen so that we expect the
inner parts of the lobes to come into pressure balance with the
external medium on large scales; we show that this leads to the
expected departure from self-similarity and the formation of
characteristic central structures in the hot external medium. The work
done by the expanding radio lobes on the external hot gas is roughly
equal to the energy stored in the lobes for all our simulations once
the lobes are well established. We show that the external pressure at
the lobe midpoint is a reasonable estimate of the internal (lobe)
pressure, with only a weak dependence on the environmental parameters:
on the other hand, the predicted radio emission from a source of a
given physical size has a comparatively strong dependence on the
environment in which the lobe resides, introducing an order of
magnitude of scatter into the jet power versus radio luminosity
relationship. X-ray surface brightness and temperature visualizations
of our simulations bear a striking resemblance to observations of some
well-studied radio galaxies.
\end{abstract}

\section{Introduction}
\label{intro}

Ever since the establishment of the standard (`beam') model for
powerful double radio sources \citep{Scheuer74,Blandford+Rees74} it has
been clear that their dynamics must depend strongly on their
environment. \cite{Scheuer74} proposed two limiting cases. In his
model A, the transverse expansion of the radio galaxy ignores the
external medium; this would be true in practise if radio galaxies were
strongly overpressured with respect to their external medium at all
times in their lifetime, in which case the lobe pressure would drive a
supersonic transverse expansion and we would expect to see an
elliptical bow shock surrounding the sources at all times. In model C,
on the other hand, the pressure in the lobes becomes comparable to the
external pressure, and then, as Scheuer put it, ``the outer parts of
the cavity swell at the expense of the parts nearest the massive
nucleus, where the thermal gas pressure is higher''. Such a model, as
Scheuer realised, is significantly harder to deal with analytically.

For this reason popular analytic models of radio lobe dynamics have
tended to be descendants of Scheuer's model A.
\cite{Begelman+Cioffi89} showed that it was possible to construct
self-consistent models in which the lobes remained overpressured, and
therefore supersonically expanding, throughout a significant part of
their lifetimes, while allowing the jet to be confined by the material
inside the cocoon. Models of this sort also provide the basis for the
important and widely used work of \cite{Kaiser+Alexander97} (hereafter
KA) who constructed an analytic model for the growth of a radio source
in a power-law atmosphere using the assumption of self-similar
expansion, which restricts the applicability of the model of the model
to the strongly overpressured phase. More recently,
\cite{Hardcastle+Worrall00} showed that the minimum pressures in radio
lobes were typically comparable to, or less than, the external thermal
pressures in the centres of the host groups or clusters, and argued
that the KA model, applied to observed extended 3CRR classical double
(\citealt{Fanaroff+Riley74} class II, hereafter FRII) sources in group
or cluster environments, required much lower cluster temperatures for
self-consistency than are actually plausible. Since then, the routine
use of X-ray inverse-Compton pressure measurements combined with the
information provided by X-ray observations on the external medium
\citep[e.g.][]{Hardcastle+02,Croston+04} have shown that our best
estimates of the lobe pressures are comparable to the external
pressure on scales comparable to those of the midpoints of the lobes,
and therefore in general significantly less than the pressure at the
centre of the cluster or group environment; either these best
estimates of pressure are wrong (for example, because the
contributions of non-radiating particles to the lobe pressure has been
severely underestimated), or lobes are {\it not} strongly
overpressured with respect to their environments, and models such as
those of KA cannot consistently describe them on the largest scales.
Analytical calculations of the structure of FRII radio sources
approaching pressure equilibrium have been presented by
\cite{Alexander02}; these show that in atmospheres in which
pressure/density decreases steeply with radius, a dense layer of
ambient gas is predicted next to the radio lobes. Its gravity squeezes
the lobes and causes the onset of non-self-similar evolution.

While analytical work is enlightening regarding the general structure
and basic physics of those sources, it relies on many assumptions,
e.g. about the geometry of the sources. Numerical modelling, in
principle, frees us of the limiting assumptions of analytic models and
allows us to model radio galaxies whose behaviour is in better
agreement with observation. However, the high-resolution simulations
required, with their large demands on simulation time, have typically
required some compromises. Early numerical modelling
\citep[e.g.][]{Norman+82,Koessl+Mueller88,Lind+89} assumed a
uniform-density environment, which is quite possibly not a bad
assumption for the early phases of a radio galaxy's growth but
certainly not valid on scales of tens to hundreds of kpc. This was
followed by 2D and 3D hydrodynamical modelling of sources in
semi-realistic ($\beta$-model) environments
\citep[e.g.][]{Reynolds+02,Basson+Alexander03,Zanni+03,Krause05}.
These generally showed deviations from self-similarity in the sense
that the axial ratios of the simulated sources did not remain
constant, but instead grew with time. More recently,
\cite*{Gaibler+09} have demonstrated in a magnetized jet simulation
that self-similarity breaks down when a radio source gets close to
pressure equilibrium. Qualitatively, similar results are seen in other
MHD simulations with realistic environments \citep[e.g.][]{ONeill+05,
  ONeillJones10}; environments derived from simulations of dynamically
active clusters are used \citep[e.g.][]{Heinz+06} and the culmination
of this type of work is represented by, for example,
\cite*{Mendygral+12} whose simulations are fully 3D, contain magnetic
fields, take account of electron transport and loss processes, and
embed the radio galaxy in an environment that is itself extracted from
a cosmological simulation. Because of the limitations in CPU time,
though, the vast majority of the work carried out in this area has
involved small numbers of simulations considering one or at most a few
different environments. It is thus difficult to get an overview of the
effects that the known range of environments, even simply considering
such basic properties as scale lengths and large-radius power-law
indices, has on the dynamics of the radio sources.

Comparison between analytical and numerical work has also given rise
to some puzzling results. As noted above, KA presented convincing
physical arguments for the existence of a self-similar expansion law
in the overpressured phase. Yet this self-similar expansion is not
easily seen in simulations \citep[e.g.][]{Carvalho+O'Dea02}. This
difficulty is related to the fact that simulations often use a jet
that is already collimated as a boundary condition to the simulation.
But, as KA showed, the self-similar expansion phase is linked in a
crucial way to the self-collimation of the jet by the overpressured
lobes (compare also model B of \citealt{Scheuer74}). Thus, if a jet is
injected into a computational domain in a parallel, collimated way,
the jet radius cannot adapt in the proper way, which results in an
expansion law that is in detail unphysical. For FRII jet simulations,
\cite{Komissarov+Falle98} have shown that a self-similar scaling is
indeed achieved if the jet is injected conically. They were even able
to account in detail for the shock structure observed in the Cygnus A
jet, produced by the self-collimation, albeit with a worryingly large
initial opening angle. \cite{Krause+12} have re-analysed the
problem using similar simulations, and found that the problem was
related to a matter of definitions: when the same definitions are
applied to simulations and observations, they are entirely consistent.
In the present paper, we aim to set up FRII radio sources where the
jet is self-collimated, so that self-similar expansion is in principle
possible, and then to investigate their behaviour when they come into
pressure equilibrium.

The desire to know exactly how radio galaxies evolve in realistic
environments is not a purely abstract one. Radio galaxies in general
provide a unique way to couple the high power of an active nucleus
(AGN) to its environment on scales of hundreds of kpc to Mpc, where
the effects of AGN radiation are expected to be negligible. They
therefore play an important role in models of galaxy formation and
evolution, where they provide so-called `feedback', helping to prevent
the cooling of large-scale gas and the consequent growth of the host
galaxies \citep[e.g.][]{Croton+06}. However, whether the physics of
radio galaxies is consistent with the role they are required to play
in these models is still an open question \citep[see e.g.][for a
  recent review]{McNamara+Nulsen12}. For example, it is not remotely
clear whether the low-power, Fanaroff-Riley class I (FRI) radio
galaxies with extended twin jets, seen in many rich group and poor
cluster environments, are having any significant energetic impact on
the gas with short cooling times, which exists on scales of a few kpc
from the nucleus \citep[e.g.][]{Hardcastle+02-2,Jetha+07}. Such
systems are very difficult to model in detail numerically (though see,
e.g., the work of \citealt{Perucho+Marti07}). Classical double radio
galaxies, although rarer, are important in this context simply because
they are required to have a strong effect on the external medium
through the driving of strong shocks, which increases the entropy of
the external medium and offsets cooling, and does so through a
comparatively large volume of the environment (much larger than the
observed radio lobes); they are also simpler to model. Indeed, some of
the early work on numerical modelling with realistic environments
\citep[e.g.][]{Basson+Alexander03,Krause05,Zanni+05,ONeill+05} was
done with the explicit aim of addressing questions about the energetic
effects of a powerful radio source. In this context, \cite{Zanni+05}
have shown that energy input by jets may reconcile cluster atmospheres
from cosmological simulations with observed X-ray scaling relations
for groups and clusters of galaxies. However, again, the small numbers
of simulations carried out in these studies mean that it is hard to
get an overview of the differences in energy transfer efficiency, if
any, that might be expected for different environments. Consequently
observers, rather than making use of the results of numerical
modelling, have a tendency to use naive estimates of the work done on
the external medium (e.g. $p{\rm d}V$ for some pressure $p$) together
with estimates of the jet power based on analytical rather than
numerical work.

In the present paper we try to address these disconnections between
numerical modelling and observation. We use the very large amount of
computational power that has become available over the course of the
past decade, not to include still more physics in a simulation of an
individual radio galaxy, but to carry out simple (though
high-resolution and large-volume) simulations of radio galaxies in a
wide range of environments. Our objective in doing so is to draw
conclusions about the dynamics and energetic input of radio galaxies
that may be generally applicable, and to test the validity of some
assumptions commonly used by observers.

\section{The simulation setup}
\label{s-setup}

Our modelling in this paper uses the freely available code
PLUTO\footnote{http://plutocode.ph.unito.it/}, version 3.1.1,
described by \cite{Mignone+07}. We chose to use PLUTO because of its
good handling of high Mach number flows and the simplicity with which
new problems can be defined. In particular, we employed Pluto's
`two-shock', exact Riemann solver, third order interpolation with the
van Leer flux limiter and third-order Runge-Kutta time-stepping. The
global accuracy of the method is, however, second order due to the way
the fluxes are calculated.

Our general approach follows that of \cite{Krause+12}; that is, we
inject two, oppositely directed, initially conical flows with high
Mach number at the centre of the simulation, and allow them to
recollimate in the external pressure. In a uniform-density
environment, we would expect this recollimation to take place on a physical
scale comparable to the inner scale defined by \cite{Alexander06},
\[
L_1 = 2\sqrt{2}\left(\frac{Q_0}{\rho_x v_j^3}\right)^{1/2}
\]
where $Q_0$ is the jet power, $\rho_x$ the external density and $v_j$
the jet speed. In such an environment, we would also expect the lobes
to come into pressure balance on scales comparable to \citep{Komissarov+Falle98}
\[
L_2 = \left(\frac{Q_0}{\rho_x c_x^3}\right)^{1/2}
\]
where $c_x$ is the external sound speed. As $L_2/L_1 = ({\cal
  M}/2)^{3/2}$, where $\cal M$ is the Mach number of the jet, it is
immediately clear that high-resolution simulations are necessary to
model high-Mach-number (and therefore realistic) jets out to the scale
on which the lobes are likely to come close to pressure equilibrium;
the cell size of the simulation needs to be $\ll L_1$ and the outer
radius of the simulation needs to be of order a few times $L_2$. For
this reason, we chose to reduce the complexity of the problem by
carrying out 2D (except as described below), axisymmetric,
hydrodynamic simulations. The co-ordinates are spherical polars and
the jets are implemented as a boundary condition at the inner radius,
injecting a conical flow with opening angle of 15$^\circ$ [chosen to
  ensure FRII-like structure: see \cite{Krause+12} for a discussion of
  the effect of opening angle on the lobe morphology]; that is, at $r
= r_{\rm inner}$, for all points with $\theta \le 15^\circ$ or $\theta
\ge 345^\circ$, we set $\rho = \rho_0$, $v_r = {\cal M}c_s$, $v_\theta
= 0$, while for all other values of $\theta$ we use a reflection
boundary condition at $r_{\rm inner}$.. A conserved tracer quantity is
injected with the jets and we use non-zero values of this quantity to
define the lobes in the simulations, as described in more detail
below.

An important feature of our modelling is that the
environments are {\it not} uniform. Instead, we represent the
environment by a standard isothermal $\beta$ model (King profile),
\[
n = n_0\left[1+\left(\frac{r}{r_{\rm c}}\right)^2\right]^{-3\beta/2}
\]
This introduces a new scale, $r_{\rm c}$ (the core radius) into the problem,
meaning that self-similar analytic approaches (e.g. KA) are no longer applicable, but it allows us to
represent a range of realistic environments from groups to clusters of
galaxies. We impose a gravitational potential, as described by \cite{Krause05}, to stabilize the $\beta$-model atmosphere.

Within the simulations, we take the unit of length to be $L_1$, the
unit of density to be $\rho_0 = n_0 \mu m_p$, where $\mu$ is the mean
mass per particle, and the unit of speed to be $c_s$, the speed of
sound in the undisturbed ambient medium. In these units, the central
pressure $p_0$ is $1/\gamma$, or 3/5. In order to choose sensible
values for $r_{\rm c}$ in these units, we need to decide on an external
calibration for these quantities. Since our interest is in low-power
FRII radio galaxies in rich group/poor cluster environments, we adopt
$Q = 10^{38}$ W ($10^{45}$ erg s$^{-1}$), a plausible value for a
low-power FRII jet \citep[e.g.][]{Daly+12}. For the basic
environmental properties we use $kT = 2$ keV, and $n_0 = 3 \times
10^4$ m$^{-3}$, which are reasonable values for the type of
environment we aim to study; we take $\mu = 0.6$. With these values we
have $c_s = 730$ km s$^{-1}$, and $L_2 \sim 100$ kpc; as the core
radii of groups or clusters are of the order of tens to hundreds of
kpc, this means that the scales that we will be studying are not too
disparate. The choice of a reasonable Mach number is then set by the
requirement that $L_2/L_1$ should not be too large for simulations at
reasonable resolution. This requires a compromise regarding the Mach
number. In reality, the Mach number should be (at least) several
hundred, which leads to a scale separation of $L_2/L_1 > 1000$. At
this point, we are only able to simulate jets out to several hundred
$L_1$, requiring ${\cal M}\la 50$ to allow us to reach scales of a few
times $L_2$. In what follows we adopt a range of Mach numbers, but our
fiducial value is ${\cal M} = 25$, which for the parameters above
gives $L_1 = 2.1$ kpc. Our typical simulation has a resolution of 2000
($r$) $\times 1600$ ($\theta$) and an outer radius of $150L_1$, which
for the parameters above is $\sim 3L_2 \approx 300$ kpc; thus we can
both sample on scales significantly less than $L_1$ (which is
necessary for the simulations to produce a self-collimating jet) and
go out to scales significantly larger than $L_2$. For these
parameters, the simulation time unit is $\tau = L_1/c_s = 2.9 \times
10^6$ years, and we expect to simulate the growth of the radio source
out to the edge of the computational volume, which (since we expect
the radio source growth to be and remain supersonic) implies
simulations lasting at least several tens of simulation time units, or
of the order $10^8$ years; this is comparable to plausible maximum
ages for real radio galaxies [e.g. to the maximum spectral ages
inferred for giant radio galaxies by \cite{Jamrozy+08}; note that
spectral ages tend to underestimate the true dynamical age \citep{Eilek96}].

Our choices of $\beta$ and $r_{\rm c}$ for the modelled atmosphere
should then reflect the variety of environments observed in real
groups or clusters. Observationally, $\beta$ varies between about 0.3
and 1.0 in groups hosting radio galaxies \citep[see,
  e.g.,][]{Croston+08}; note that we are only simulating the
large-scale environment and so comparisons should be made with $\beta$
rather than $\beta_{\rm in}$ in this paper -- the small-scale dense
cores seen in observations would only affect the dynamics of the radio
galaxy over a comparatively short period). Core radii span the range
of tens to hundreds of kpc, as noted above. As described in more
detail in the next section, we carry out simulations using a range of
$\beta$ and core radius values to allow us to investigate which
properties of the simulated radio galaxy are dependent on, and which
are independent of, the choice of environment.

We choose to model bipolar jets because this avoids any uncertainty in
setting the correct boundary conditions on the plane $\theta=\pi/2$
radians (the plane passing through the origin and perpendicular to the
jet axis). To break the symmetry, we impose a slight sinusoidal
variation on the Mach number of the two jets, such that ${\cal M} =
{\cal M}_0 [1+ A \sin(\omega t + \phi)]$, where $t$ is the simulation
time in internal units, $\omega$ and $A$ are constants ($A\ll 1$), but
$\phi$ is different for the two jets. As the rate of energy input into
the simulations goes as ${\cal M}^3$, this in principle affects the
calculation of the appropriate length scales given above, but the
time-averaged energy input scaling factor is $1+\frac{3}{2}A^2$, so
for small $A$ the effect is negligible. In all our simulations we used
$\omega = 2$, $A = 0.1$; $\phi=0$ for the jet propagating in the
$\theta = 0$ direction and $1$ for the jet in the $\theta = \pi$
direction. This process has the advantage that we have two
quasi-independent jet simulations for each run we carry out. The
results of these may be presented separately to assess the scatter in
any measured quantity, or may be averaged to reduce it.

Important aspects of the astrophysics of the interactions of radio
galaxies are omitted by this simulation setup. For example, there is
no radiative cooling, either in the external gas or in the material
inside the lobes. There is no particle acceleration; the material
injected in the jets is, and remains, ordinary gas with a
non-relativistic equation of state. There are no bulk relativistic
motions (for example, ${\cal M} = 25$ corresponds to $0.06c$ for the
parameters described above). Magnetic fields are not simulated, and,
of course, the simulations are axisymmetric and two-dimensional. Our
hope is, however, that the global dynamical properties of the sources
and their energetic impact should be largely unaffected by these
simplifications, which mainly affect the detailed behaviour of the
plasma inside the lobes. We will discuss the effects that these
limitations have on the interpretation of our results, together with
inferences that can be drawn from other authors' more sophisticated
simulations, in later sections.

\section{The simulation runs}
\label{s-runs}

\begin{table*}
\caption{The parameters of the simulation runs discussed in the paper.
  The columns give the type of simulation, as discussed in the text
  (Section \ref{s-runs}); the sidedness of the jet (1 for 3D
    runs, 2 for all others); the Mach number and the corresponding
  simulation length and timescale, as discussed in the text (Section
  \ref{s-setup}); the outer radius of the simulation, in internal units;
  the resolution; the properties of the environment; the timescale for
  the simulation to run to completion, in internal units; and the code
  used to designate the simulation in the rest of the paper. Where a
  column is blank, it has the same value as the nearest non-blank
  entry immediately above it.}
\label{runs}
\begin{tabular}{lrrrrrrrrrl}
\hline
Run type&Sided&$\cal M$&$L_1$&$\tau$&Outer
radius&Resolution&$\beta$&$r_{\rm c}$&$t_{\rm end}$&Code\\
&&&(kpc)&(Myr)&($L_1$)&(cells, $r \times \theta (\times \phi)$)&&($L_1$)&($\tau$)\\
\hline
Standard&2&25&2.1&2.9&150&$2000 \times 1600$&0.35&20&44.2&M25-35-20\\
&&&&&&&0.35&30&51.7&M25-35-30\\
&&&&&&&0.35&40&60.0&M25-35-40\\
&&&&&&&0.55&20&31.3&M25-55-20\\
&&&&&&&0.55&30&29.2&M25-55-30\\
&&&&&&&0.55&40&48.1&M25-55-40\\
&&&&&&&0.75&20&22.9&M25-75-20\\
&&&&&&&0.75&30&37.7&M25-75-30\\
&&&&&&&0.75&40&49.6&M25-75-40\\
&&&&&&&0.90&20&21.7&M25-90-20\\
&&&&&&&0.90&30&34.1&M25-90-30\\
&&&&&&&0.90&40&42.6&M25-90-40\\
Large&2&25&2.1&2.9&250&$3000 \times 2400$&0.35&40&95.0&M25-35-40-L\\
&&&&&&&0.55&40&77.1&M25-55-40-L\\
&&&&&&&0.75&40&60.1&M25-75-40-L\\
&&&&&&&0.90&40&45.0&M25-90-40-L\\
Varying $\cal M$&2&10&8.5&11.4&38&$500 \times 400$&0.75&7.59&7.6&M10-75-30\\
&&15&4.6&6.2&70&$930 \times 740$&0.75&13.94&18.2&M15-75-30\\
&&20&3.0&4.0&107&$1430 \times 1150$&0.75&21.47&18.5&M20-75-30\\
&&30&1.6&2.2&197&$2630 \times 2100$&0.75&39.44&46.7&M30-75-30\\
&&35&1.3&1.7&248&$3310 \times 2650$&0.75&49.70&47.3&M35-75-30\\
&&40&1.1&1.4&304&$4048 \times 3238$&0.75&60.71&73.9&M40-75-30\\
3D&1&25&2.1&2.9&150&$2000 \times 800 \times 20$&0.90&30&25.1&M25-90-30-3D\\
&&&2.1&2.9&100&$2000 \times 800 \times 20$&0.75&20&16.3&M25-90-30-3D\\
\hline
\end{tabular}
\end{table*}

We carried out several basic types of simulation run:
\begin{enumerate}
\item Runs with ${\cal M} = 25$ and outer radius $150L_1$ as described
  in the previous section. These were our standard runs and sampled a
  $4 \times 3$ grid in $\beta$ and $r_{\rm c}$, where $\beta$ could take the
  values 0.35, 0.55, 0.75 or 0.90 and $r_{\rm c}$ the values 20, 30 or
  40$L_1$, corresponding to 42, 63 or 84 kpc respectively.
\item Larger runs with ${\cal M} = 25$ and outer radius $250L_1$.
  These were intended to allow us to investigate the late-time
  behaviour of the sources, and correspond to total source sizes of
  $500L_1$ or $\sim 1$ Mpc, i.e. to scales comparable to the largest
  known radio galaxies. Because they required longer run times, we
  did not run these over the full grid of $\beta$ and $r_{\rm c}$ values.
\item Runs with ${\cal M} \neq 25$. These
  enable us to search for any trends with Mach number. To allow
  like-for-like comparison between these and the other runs, we scaled
  the resolution, the core radius and the outer radius so as to
  retain (1) the same resolution as a fraction of $L_1$, (2) a core
  radius in physical units that is matched to some of the ${\cal M} =
  25$ simulations, and (3) the same physical outer radius. We chose to
  run all of these simulations for a `representative' set of cluster
  parameters, $\beta = 0.75$ and $r_{\rm c} = 63$ kpc, i.e. $30L_1$ for
  ${\cal M} = 25$. For ease of comparison between simulations of
  different Mach number, all simulation lengths in plots are scaled to
  physical units (kpc) by multiplying by $L_1$ in what follows.
\item 3D runs. These were designed to check that our results are not
  simply a result of the use of 2D simulations by having a number $>1$
  of cells in the $\phi$ axis of spherical polars. This makes the
  problem computationally considerably harder, since we need to retain
  the resolution of the 2D runs in the $r$ and $\theta$ directions in
  order to sample the jet injection region appropriately, as discussed
  above. Consequently, the final 3D runs we present have comparatively
  coarse resolution in the $\phi$ axis, and are also restricted to
  one-sided jets. They are roughly matched to
  two of the 2D runs (one has a smaller outer radius) and have ${\cal
    M} = 25$. The axisymmetry of the problem is broken in a
    similar way to the approach taken with two-sided jets, by
    adding a small precessing sinusoidal perturbation of the injection
    Mach number as a function of time and $\phi$. We only use these runs in comparisons with the corresponding
2D runs.
\end{enumerate}

Table \ref{runs} summarizes the runs that we refer to in the rest of
the paper. Each run is given a short code (e.g. `M25-90-30',
corresponding to ${\cal M}=25$, $\beta = 0.9$ and $r_{\rm c} = 30$) which
indicates the key parameters of the run, and these codes will be used,
for brevity, throughout the remainder of the paper.

All our production runs were carried out using the STRI cluster of the
University of Hertfordshire, using either 128 or 192 Xeon-based cores,
typically for about 24-48 hours per run. Some early runs to explore
parameter space were carried out on smaller numbers of cores either on
the STRI cluster or on the UCL supercomputer Legion (access provided
by the Miracle project).

Post-processing was also carried out on the STRI cluster. PLUTO was
configured to write out the complete state of the simulation every
$0.1$ simulation time units, and these images of the simulation grid
were used to compute derived quantities such as the dimensions of the
lobe, the energy stored in the lobe and in the shocked region, and so
forth. These quantities are used in the following section. One crucial
choice concerns the exact definition of the lobe: we defined material
as being inside the lobe if it was within the outermost surface where
the lobe tracer quantity was $>10^{-3}$. We use a value $<1$ here to
account for mixing, which leads to dilution of the original jet
material. The exact results are insensitive to the value used as long
as it is not too close to unity, which would lead to spurious
`heating' of the external material due to mixing with lobe material, a
process which does not appear to occur in real radio galaxies, where a
sharp boundary between the lobes and the external medium is always
seen. Material is taken to be within the shocked region if it is not
in the lobe and if it is inside the outermost surface where the radial
velocity $>10^{-3}$ simulation units. Calculation of energies for the
shocked region take as their zero point at any given time $t$ the
energy stored within the boundary of the shock at time $t$ in the
image from $t=0$, so that we ignore the pre-existing internal energy
of the environment.

\section{Results}

\subsection{General properties of the simulations}
\label{general}

\begin{figure*}
\includegraphics[width=13cm]{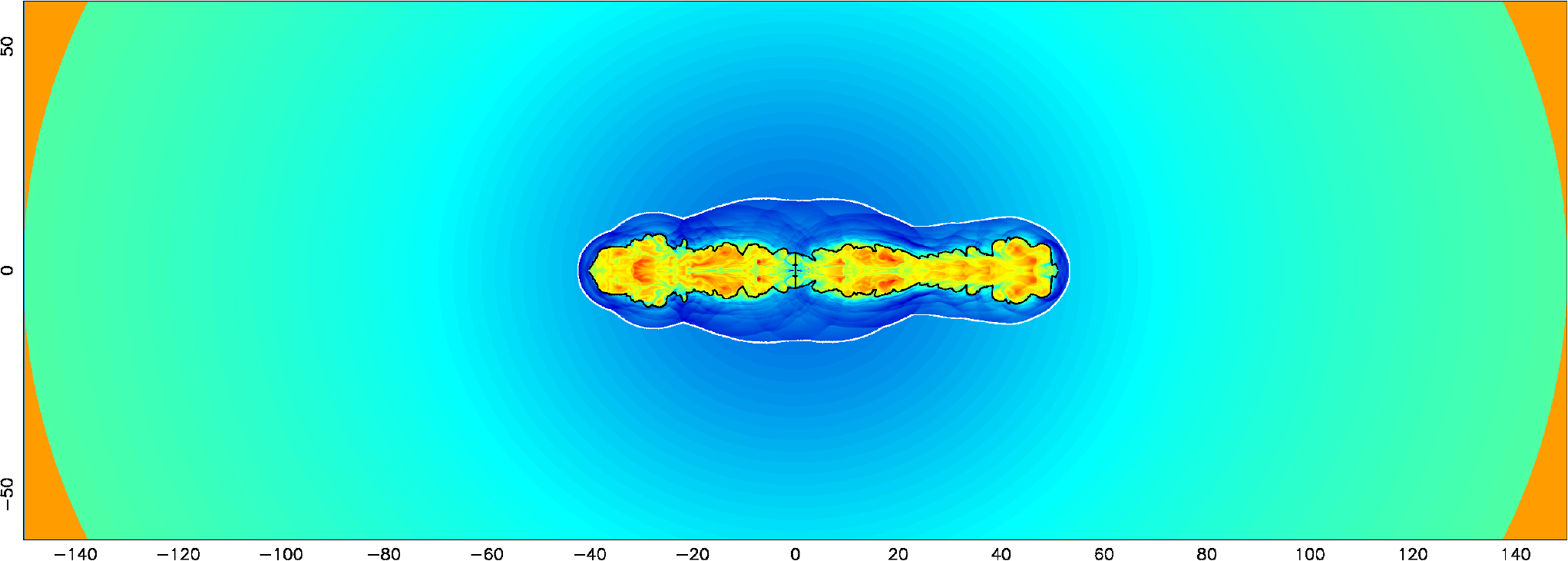}
\vskip -8pt
\includegraphics[width=13cm]{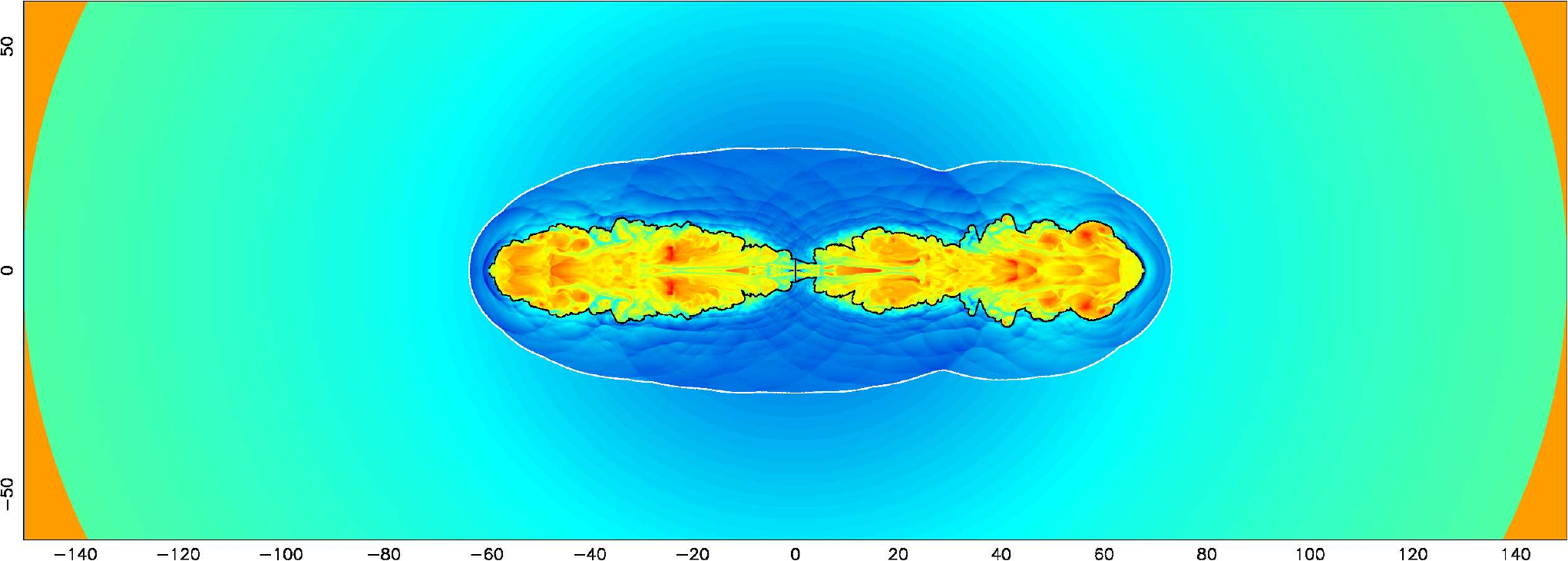}
\vskip -8pt
\includegraphics[width=13cm]{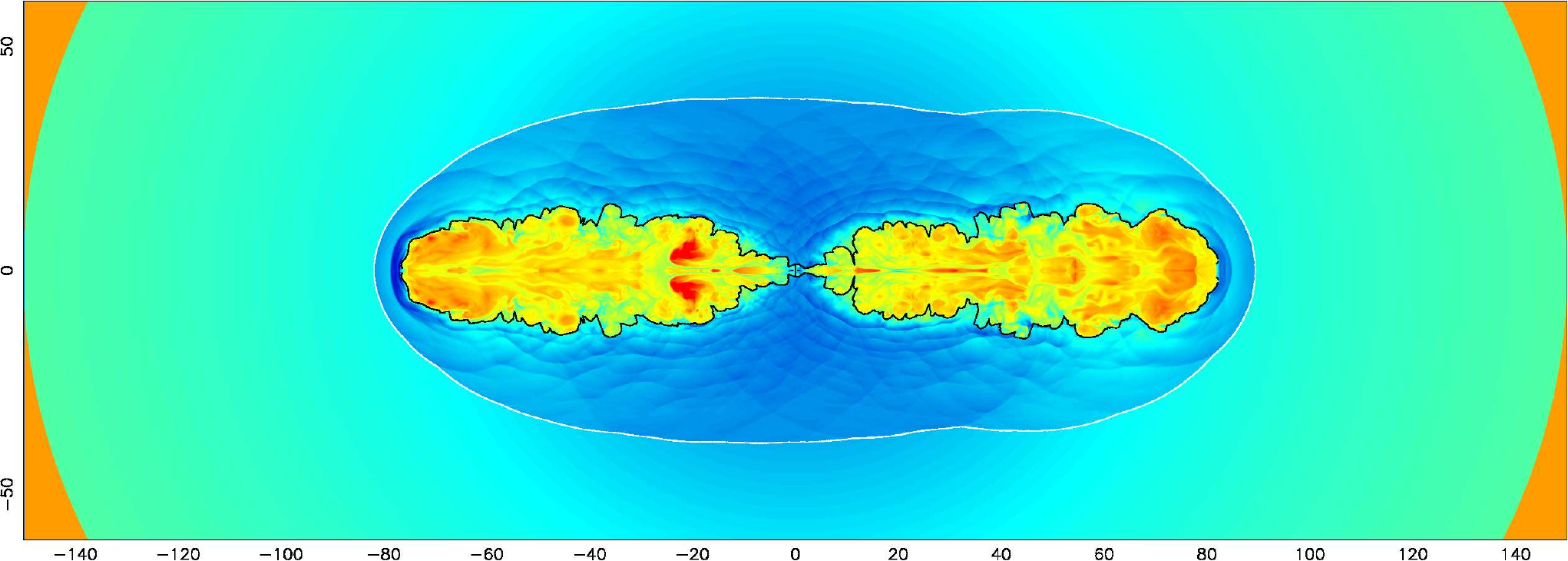}
\vskip -8pt
\includegraphics[width=13cm]{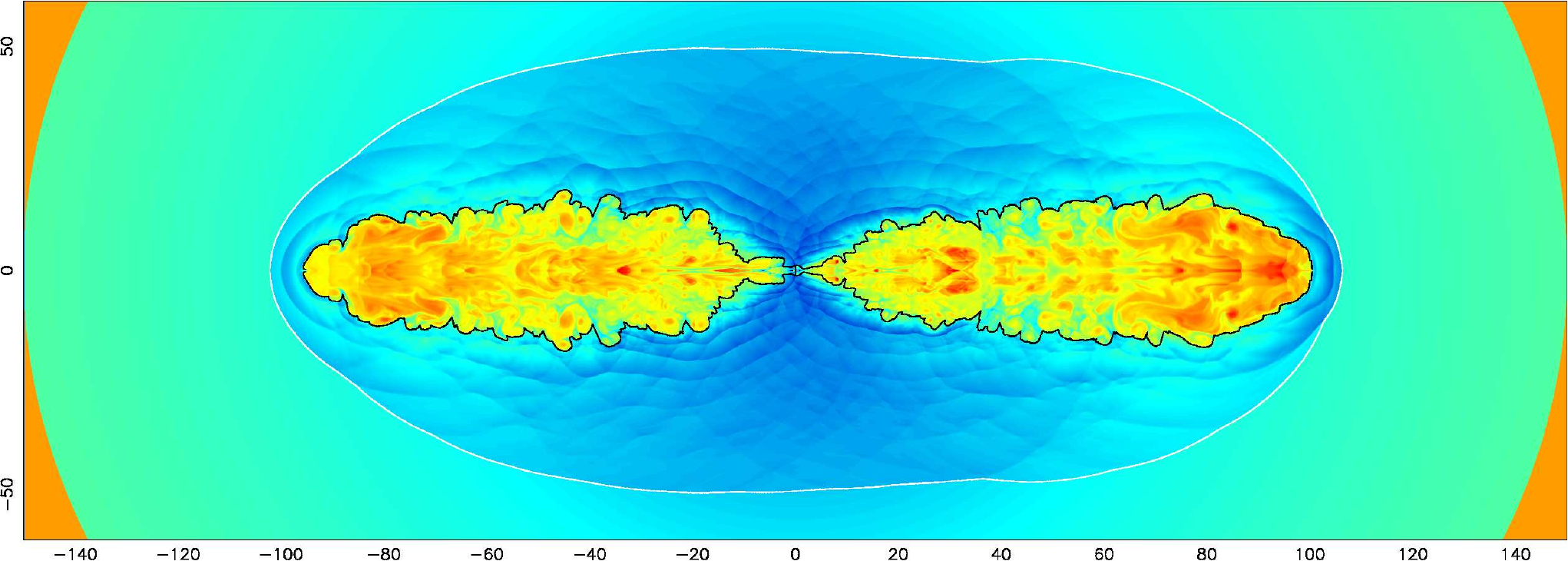}
\vskip -8pt
\includegraphics[width=13cm]{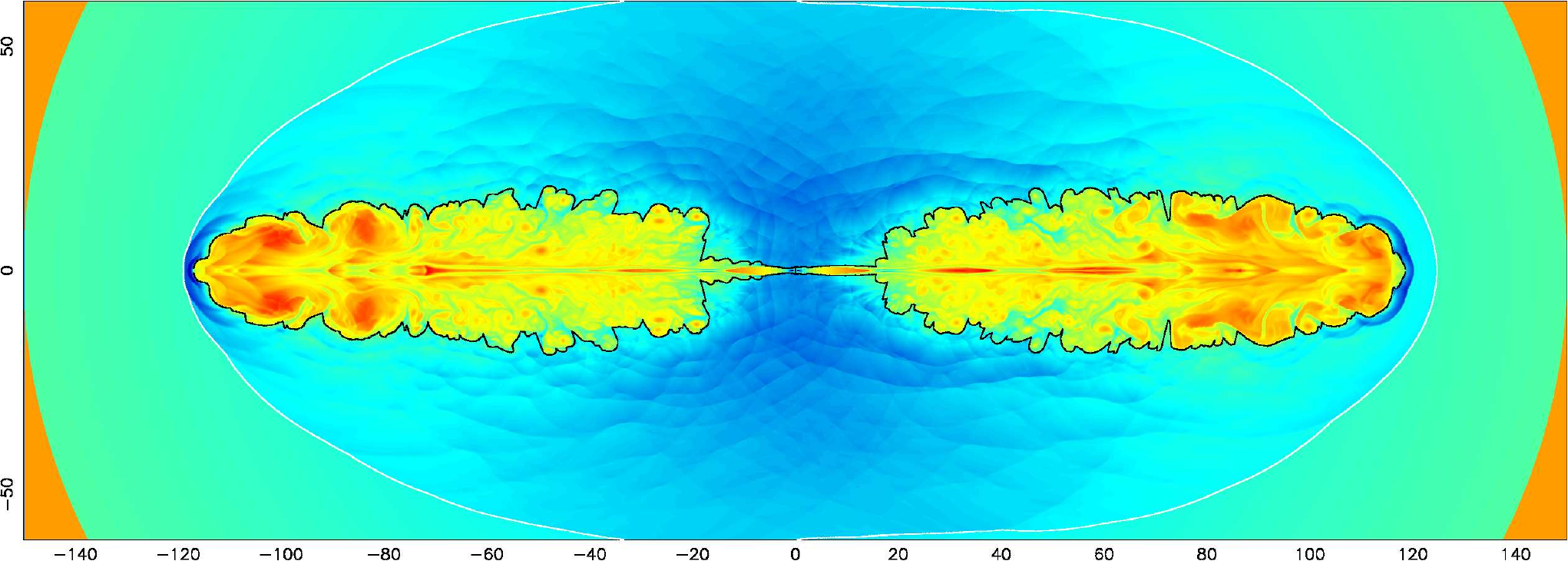}
\caption{Density in the simulation M25-35-40 for simulation times
  $t = 10$, 20, 30, 40 and 50. The images show a slice through the
  midplane of the notional 3D volume; as the simulations are
  axisymmetric, all images are reflection-symmetric about the jet
  axis. Colours show a logarithmic scale of density between 0.025 and
  6.3 simulation units, with dark blue being densest and red least dense. Black and white contours show the lobe and
  shocked region boundaries respectively, as defined in Section
  \ref{s-runs}. Units of the axes are simulation units ($L_1$).}
\label{density-example}
\end{figure*}

\begin{figure*}
\includegraphics[width=13cm]{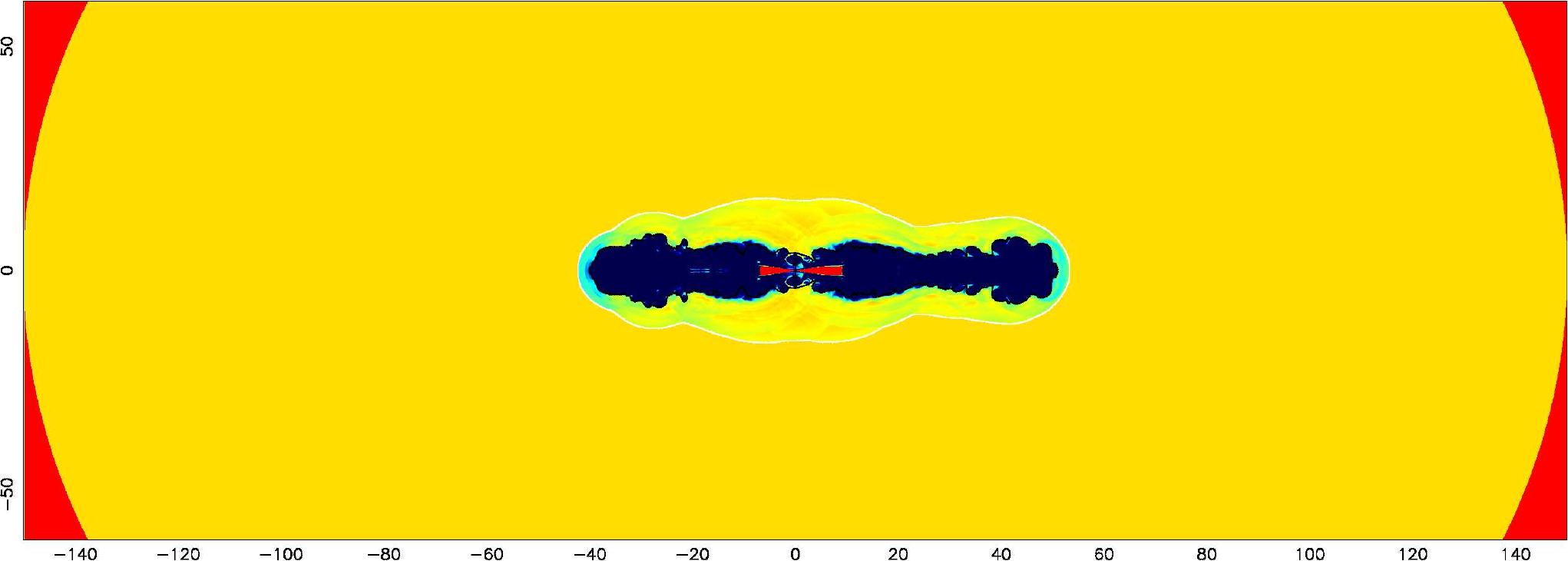}
\vskip -8pt
\includegraphics[width=13cm]{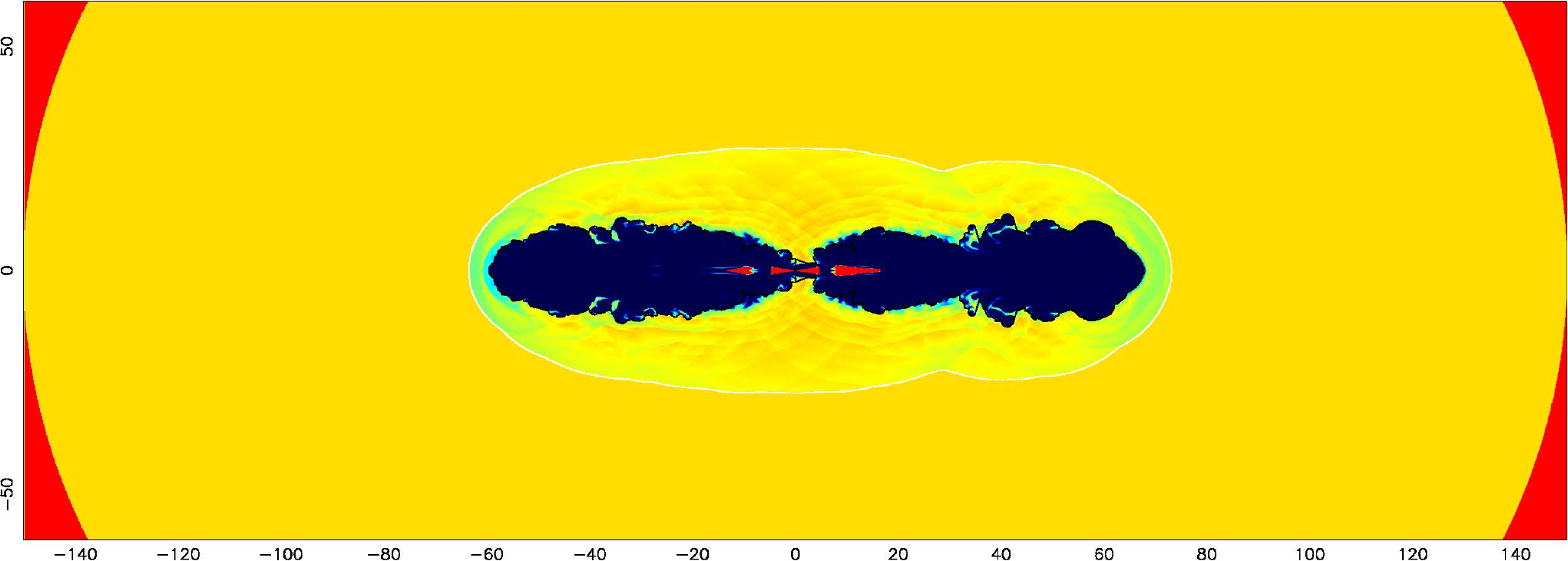}
\vskip -8pt
\includegraphics[width=13cm]{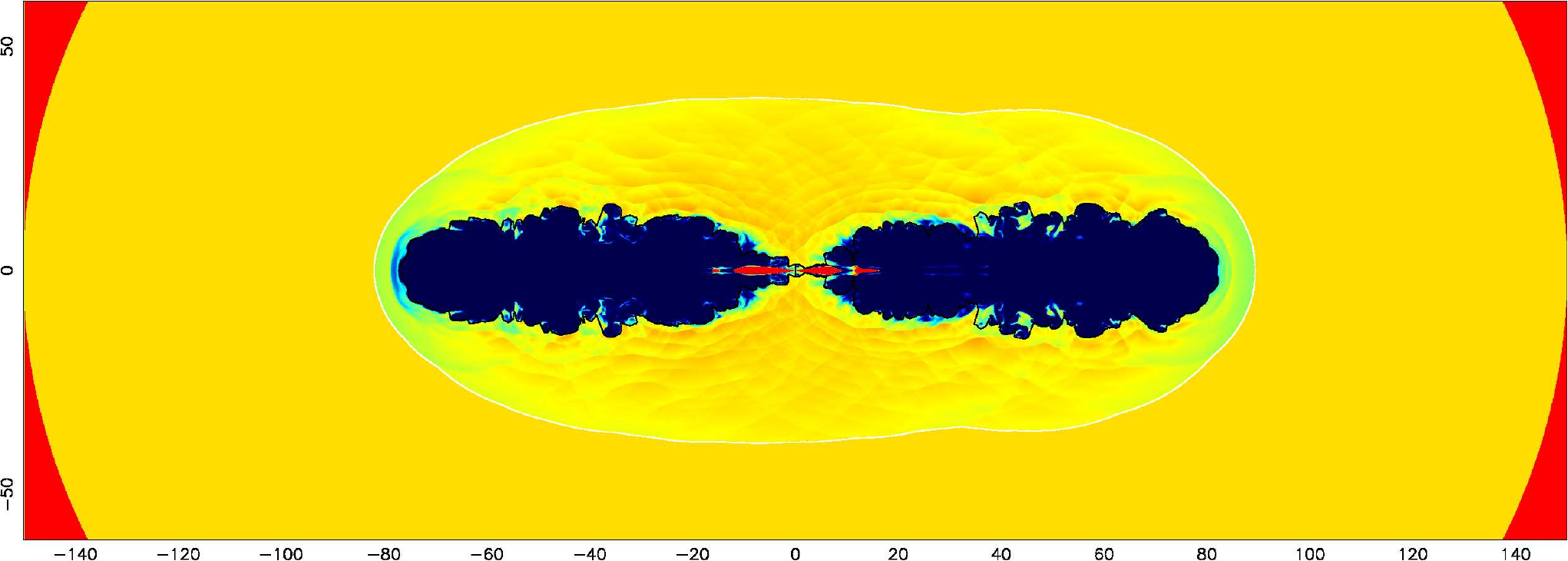}
\vskip -8pt
\includegraphics[width=13cm]{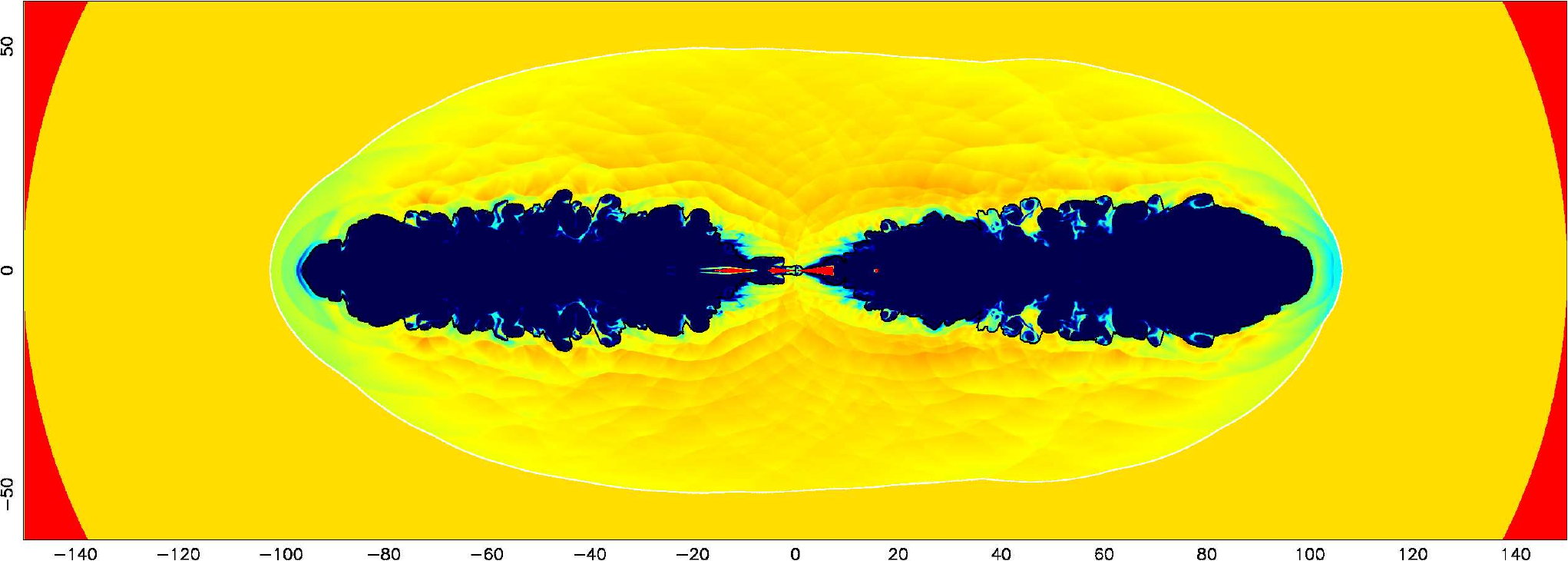}
\vskip -8pt
\includegraphics[width=13cm]{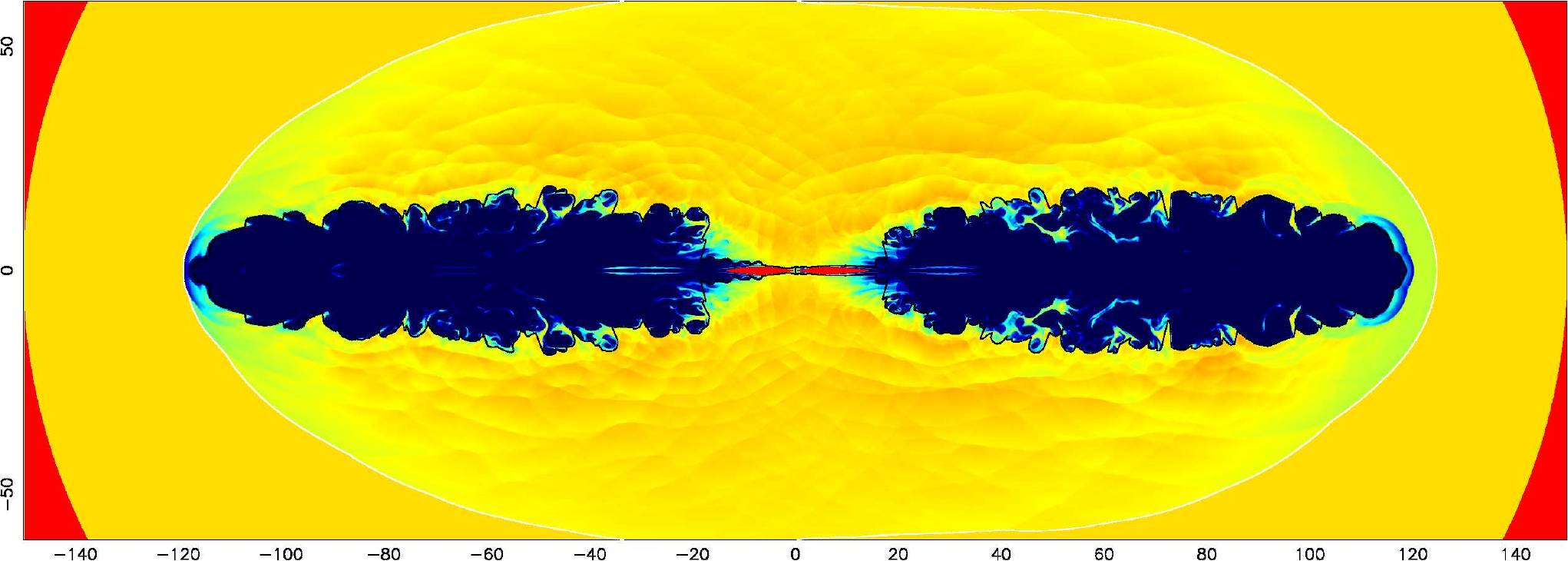}
\caption{Temperature in the simulation M25-35-40 for simulation times
  $t = 10$, 20, 30, 40 and 50. Notes as for previous figure, but
  colours show temperatures between 0.32 and 3.2 simulation units,
  with dark blue being hottest and red coldest.}
\label{temperature-example}
\end{figure*}

All simulations produce reasonably plausible `radio lobes' after the
initial establishment and collimation of a jet. Figs
\ref{density-example} and \ref{temperature-example} show maps of the
density and temperature as a function of position in Cartesian
co-ordinates for several different times in an example simulation
(M25-35-40), illustrating the general trends we observe. 

At the start of the simulations the jets initially form quite long,
thin lobes, but these then expand transversely as the material that
has travelled up the jet is thermalized by shocks driven into the
lobes by new jet material. This initial phase is a necessary
consequence of an initially overdense conically expanding jet: the
width of the lobes is related to the ratio of the jet density to the
ambient density \citep[e.g.][]{Krause03}, and only very underdense jets form
wide radio lobes. The thin lobe phase that we see is therefore a
realistic feature, but it appears on unrealistically large scales, set
by the choice of our Mach number (compare above); on these scales, it
is an artefact of our particular approach. This feature appears
repeatedly in the analysis below.

Very high temperatures are observed within the lobes at all times, as
expected given the high Mach number of the jet termination shock.
However, a stable terminal shock (corresponding to the hotspot of FRII
radio galaxies) does not develop; the jet tends to terminate soon
after entering the lobes (this can be seen in the temperature maps --
the unshocked jet is cold due to adiabatic expansion) and is unstable,
with repeated jet terminations at different positions inside the lobe
driving shocks into the lobe material itself. We do not regard this as
a major problem, because the object of our simulations is to consider
the dynamics of an overpressured lobe, rather than to reproduce the
detailed observational properties of FRII radio galaxies. The density
plots also show quite clearly that, on scales comparable to $L_2$, the
lobes start to be driven away from the central parts of the (shocked)
atmosphere. By the late stages of the simulation, there is a
substantial gap between the two lobes and the jets are propagating
through the shocked external medium without a cocoon to protect them.
Finally, we see that the edges of the lobes are Kelvin-Helmholtz
unstable, as is frequently observed in purely hydrodynamical
simulations of radio lobes: the effects of entrainment of shocked
external material (denser, colder) can be seen in the density and
temperature plots. We do not believe that this represents the
behaviour of real radio sources, noting that magnetic fields would
suppress the K-H instability \citep{Gaibler+09} but again we do not
expect the K-H instabilities to have a significant effect on the lobe
dynamics, which are governed by large-scale pressure and momentum flux
balance between the lobe and the external medium, though the
instabilities do clearly have an effect on the detailed appearance of
the simulated sources. Only if the K-H instabilities grew to size
scales comparable to those of the lobes would we expect the overall
dynamics to be affected, and this does not occur in our simulations.

Turning to the effect on the external medium, we see that initially
the lobes drive the expected shell of hot, shocked gas, though even at
early times high temperatures are only seen close to the ends of the
radio lobes. The transverse expansion of the lobe is only trans-sonic
from early times and so the post-shock gas away from the ends of the
lobes is free to expand almost adiabatically, driving at most a weak
shock into the external medium (in the example shown, we see that the
shocked gas extends to a radius of 60 simulation units in the
transverse direction at $t=50$). At late times, although there are
temperature and density fluctuations in the shocked medium (due to
weak shocks related to vortices in the lobes, which are probably not
realistic in detail), its overall temperature and density away from
the tips of the lobes is not markedly different from the initial
conditions, except of course where the lobes have formed cavities.
Note in particular the formation of a bar-like structure and arms of
{\it colder} material around the base of the lobes by $t = 50$,
structures whose visibility would presumably only be enhanced if
radiative cooling were included in our simulations. We will return to
the observational consequences of this in Section \ref{xray}.

\subsection{Dynamics of the radio lobes}
\label{dynamics}

\begin{figure*}
\includegraphics[width=8.5cm]{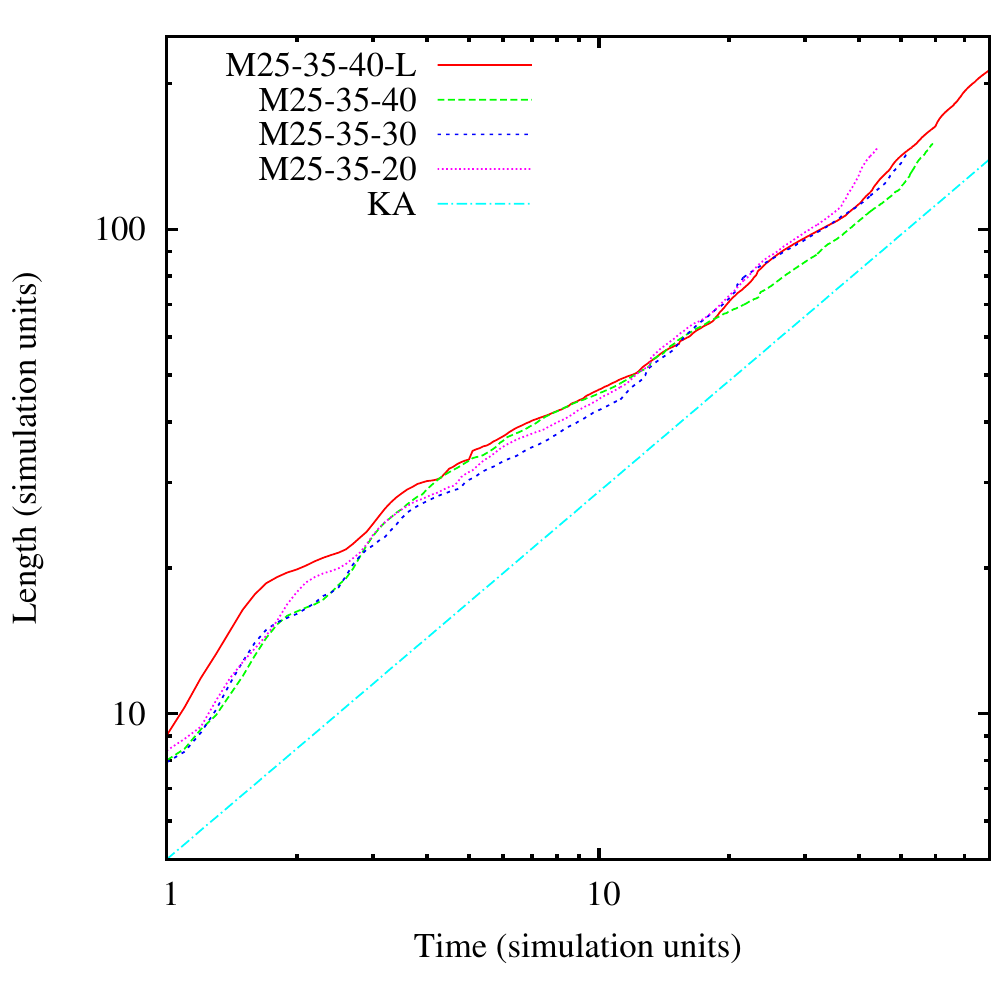}
\includegraphics[width=8.5cm]{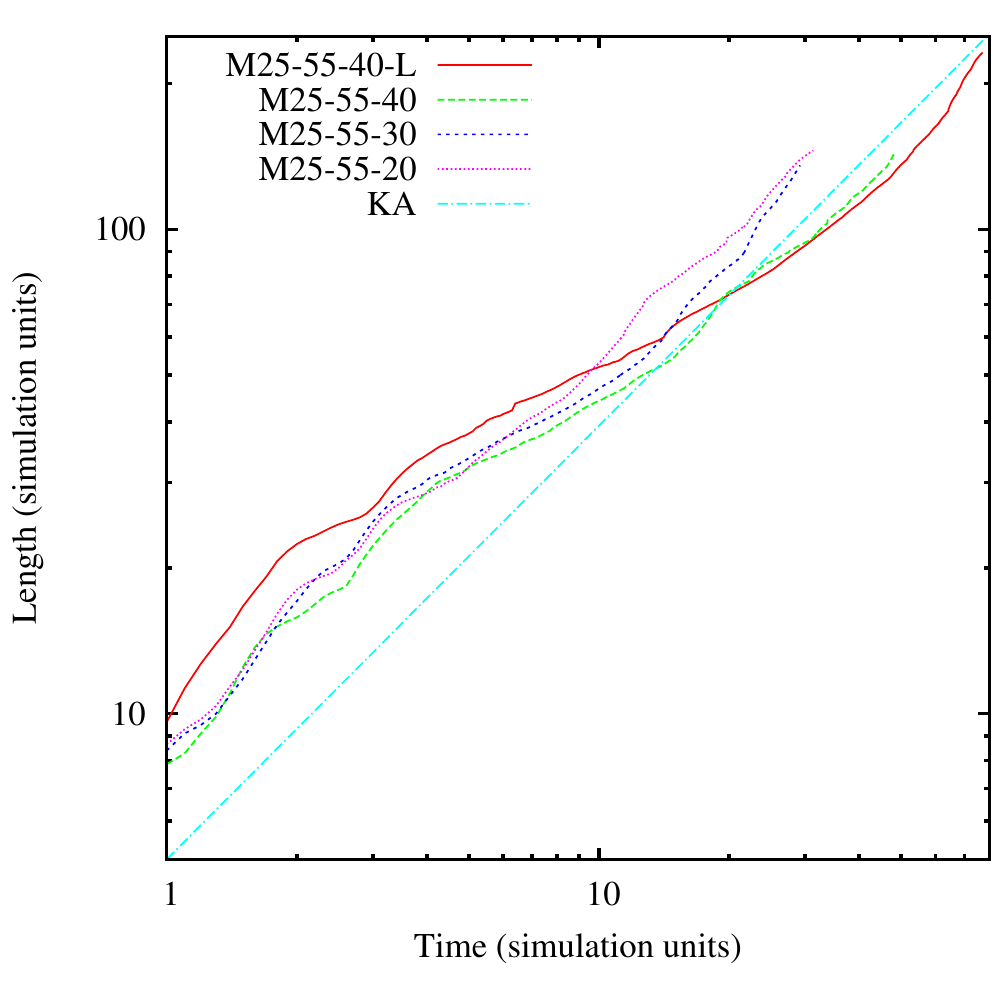}
\includegraphics[width=8.5cm]{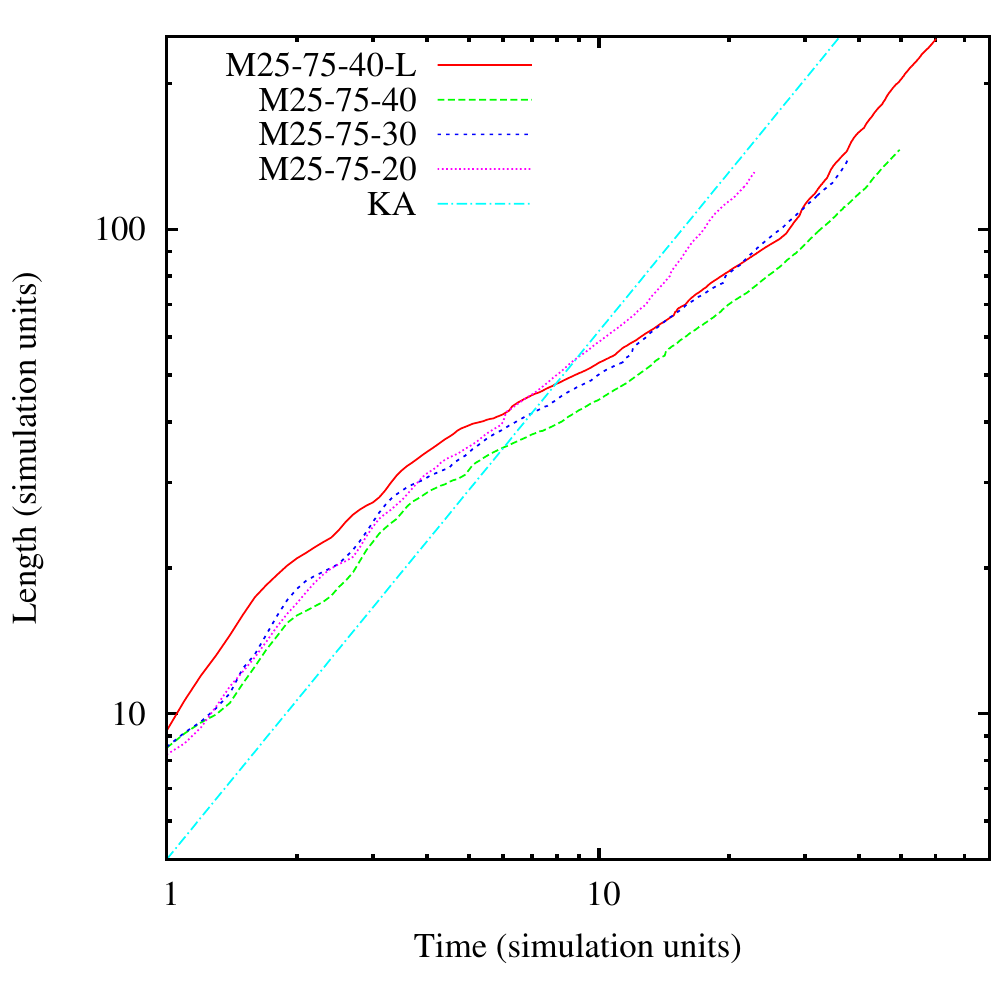}
\includegraphics[width=8.5cm]{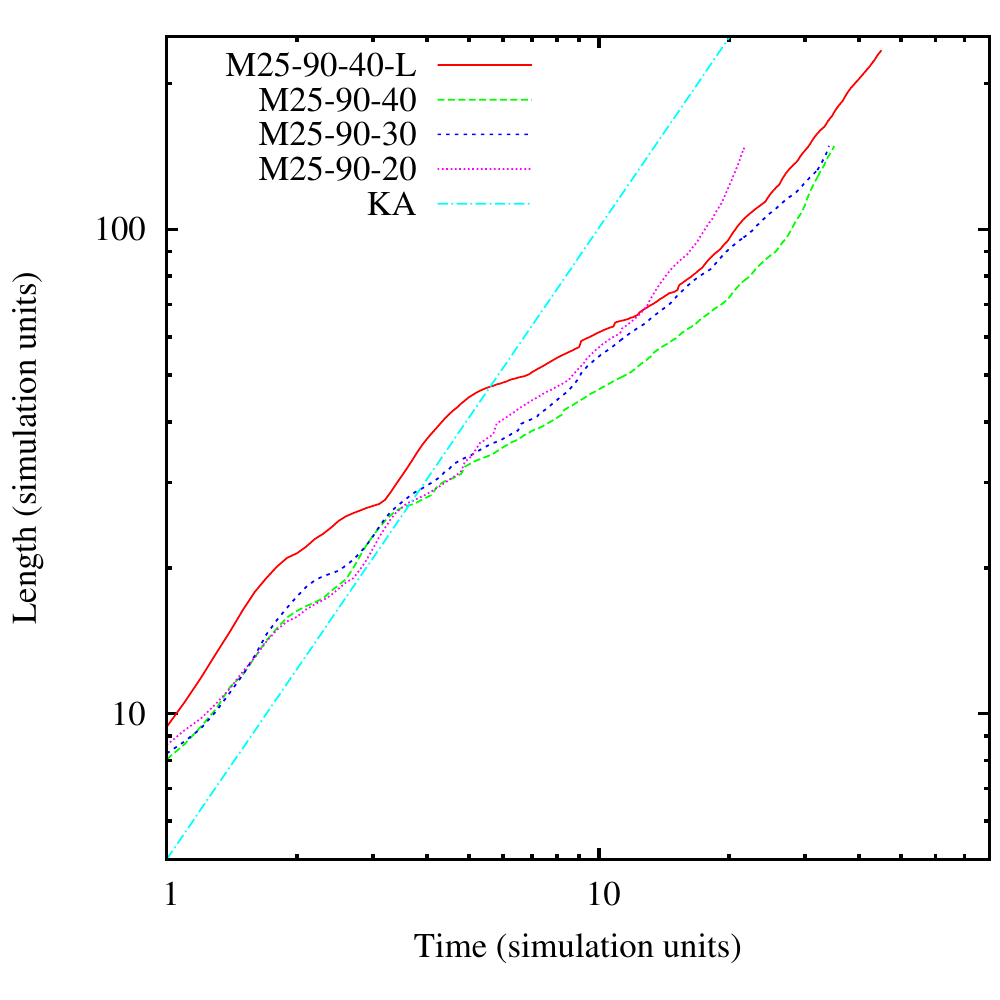}
\caption{Mean lobe length as a function of simulation time in ${\cal
    M}=25$ simulations for (top left) $\beta = 0.35$, (top right)
  $\beta = 0.55$, (bottom left) $\beta = 0.75$ and (bottom right)
  $\beta = 0.90$. Also plotted (`KA'), with arbitrary normalization,
  are the \cite{Kaiser+Alexander97} predictions for a power-law
  atmosphere.}
\label{length}
\end{figure*}

\begin{figure*}
\includegraphics[width=8.5cm]{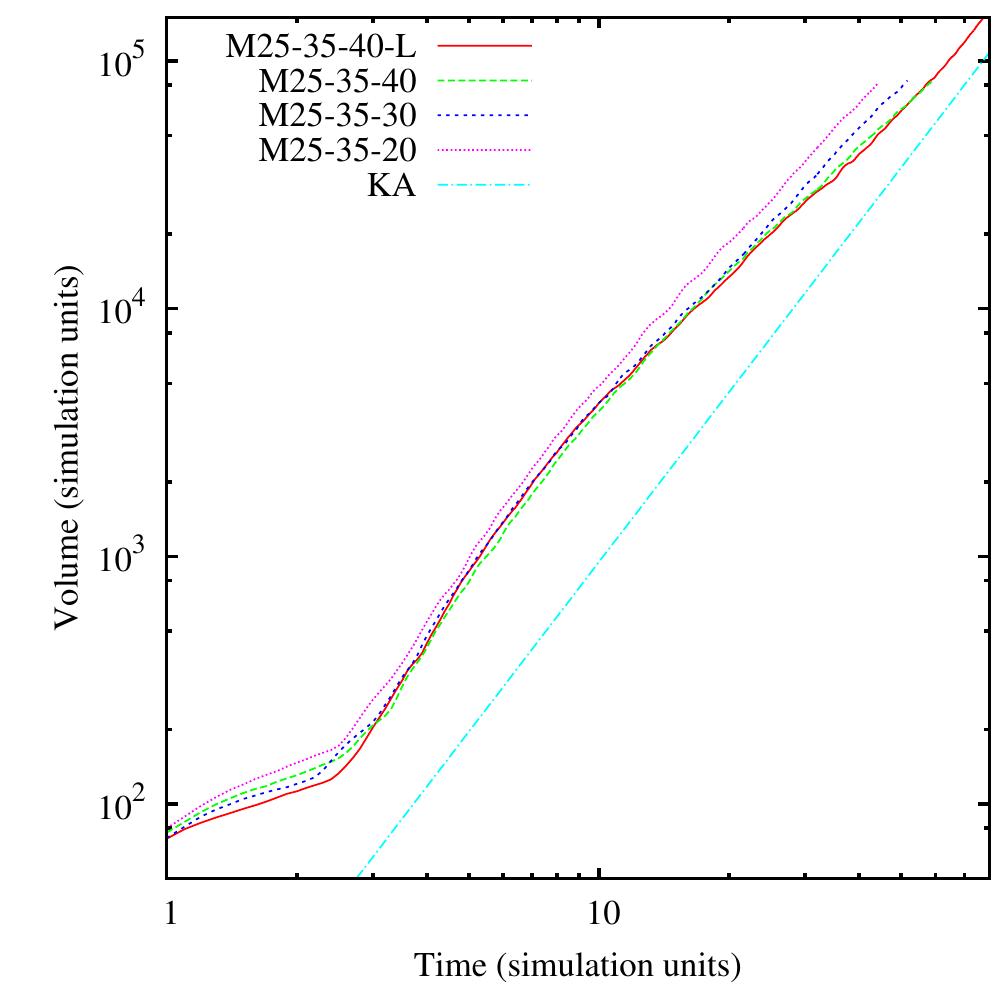}
\includegraphics[width=8.5cm]{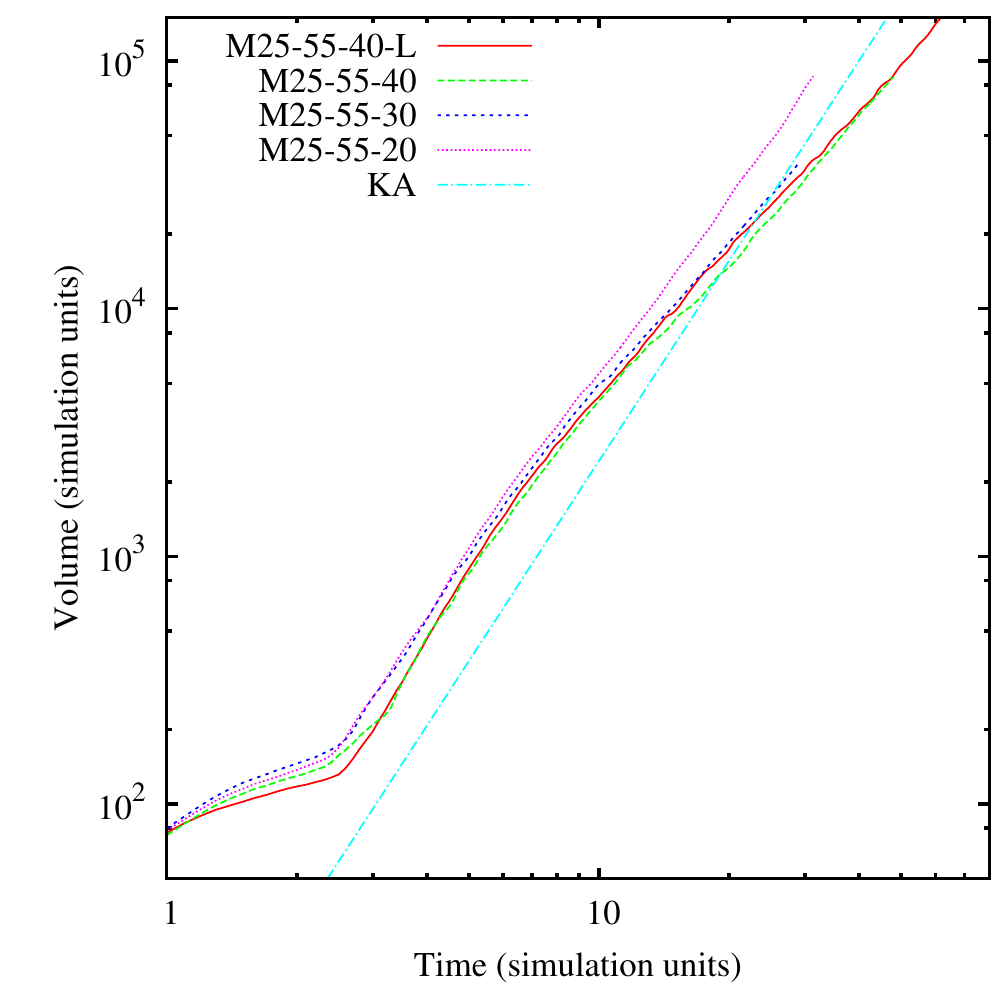}
\includegraphics[width=8.5cm]{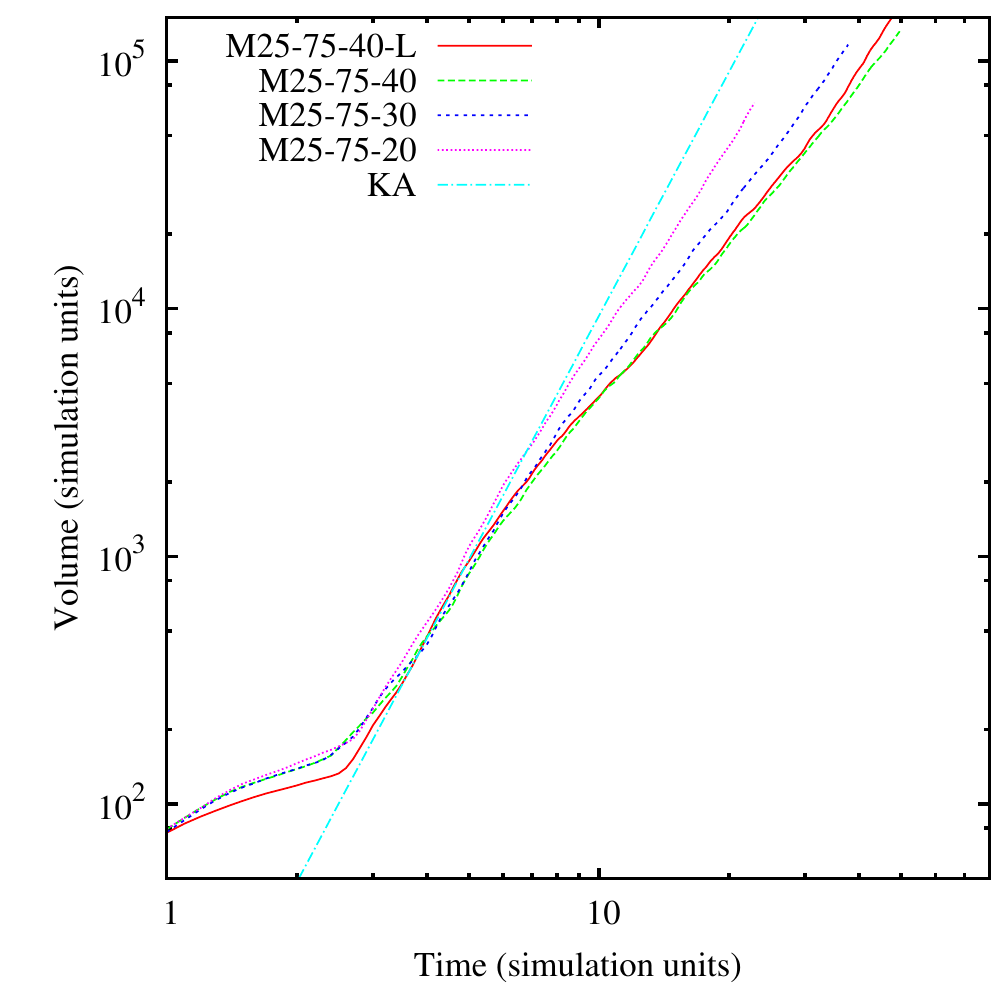}
\includegraphics[width=8.5cm]{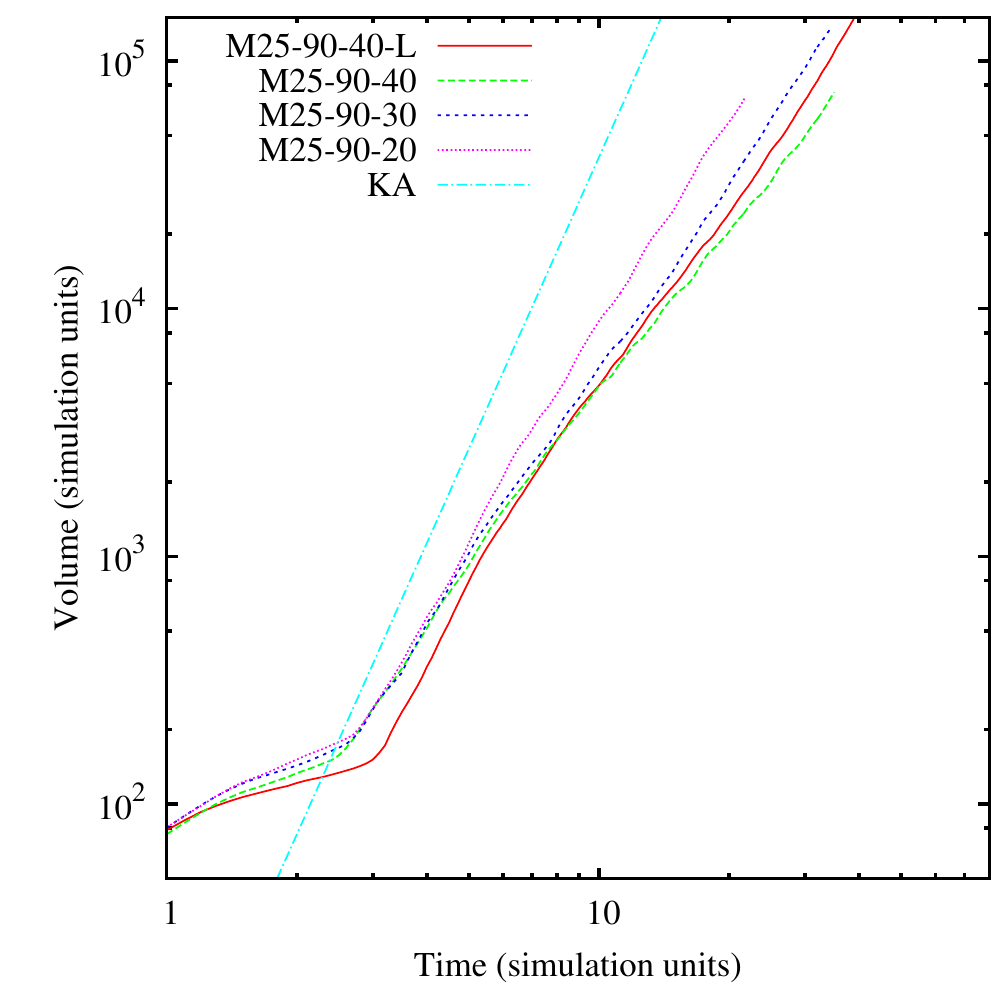}
\caption{Mean lobe volume as a function of simulation time in ${\cal
    M}=25$ simulations for (top left) $\beta = 0.35$, (top right)
  $\beta = 0.55$, (bottom left) $\beta = 0.75$ and (bottom right)
  $\beta = 0.90$. Also plotted (`KA'), with arbitrary normalization, are the \cite{Kaiser+Alexander97} predictions for a power-law atmosphere.}
\label{volume}
\end{figure*}

\begin{figure*}
\includegraphics[width=8.5cm]{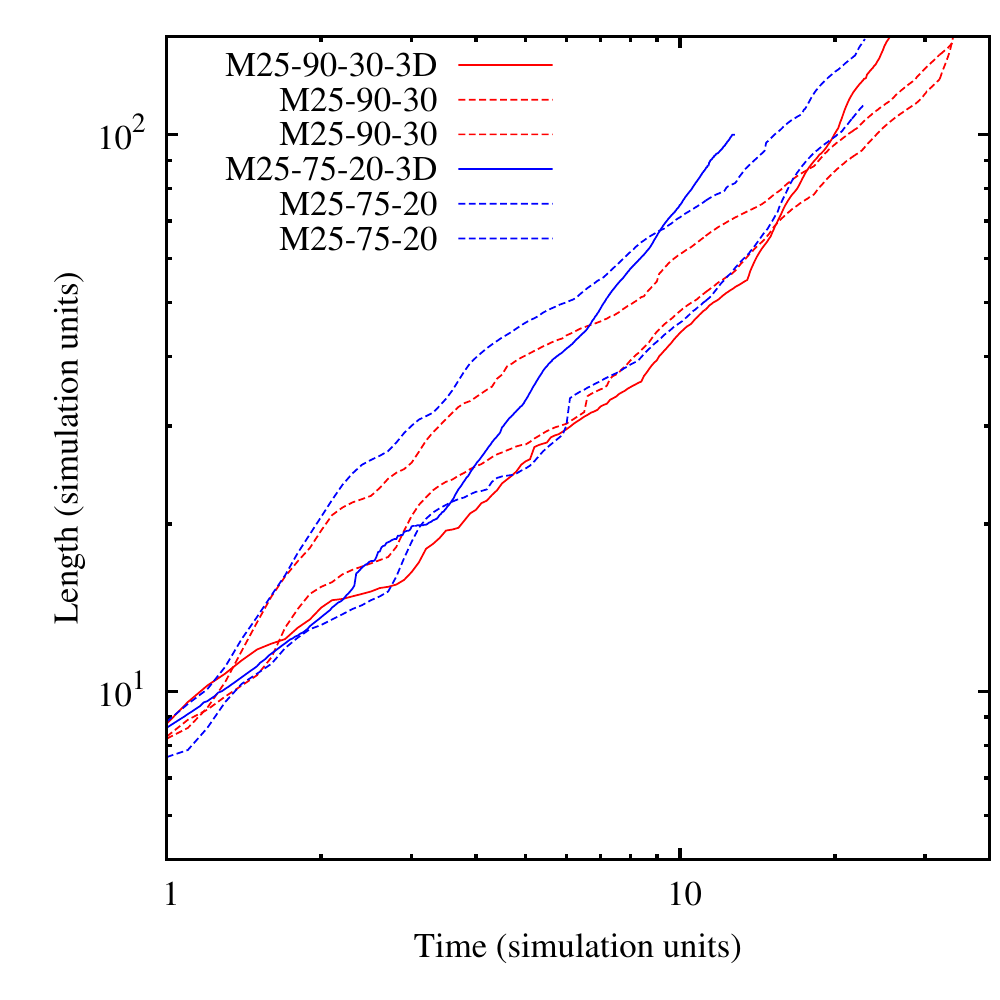}
\includegraphics[width=8.5cm]{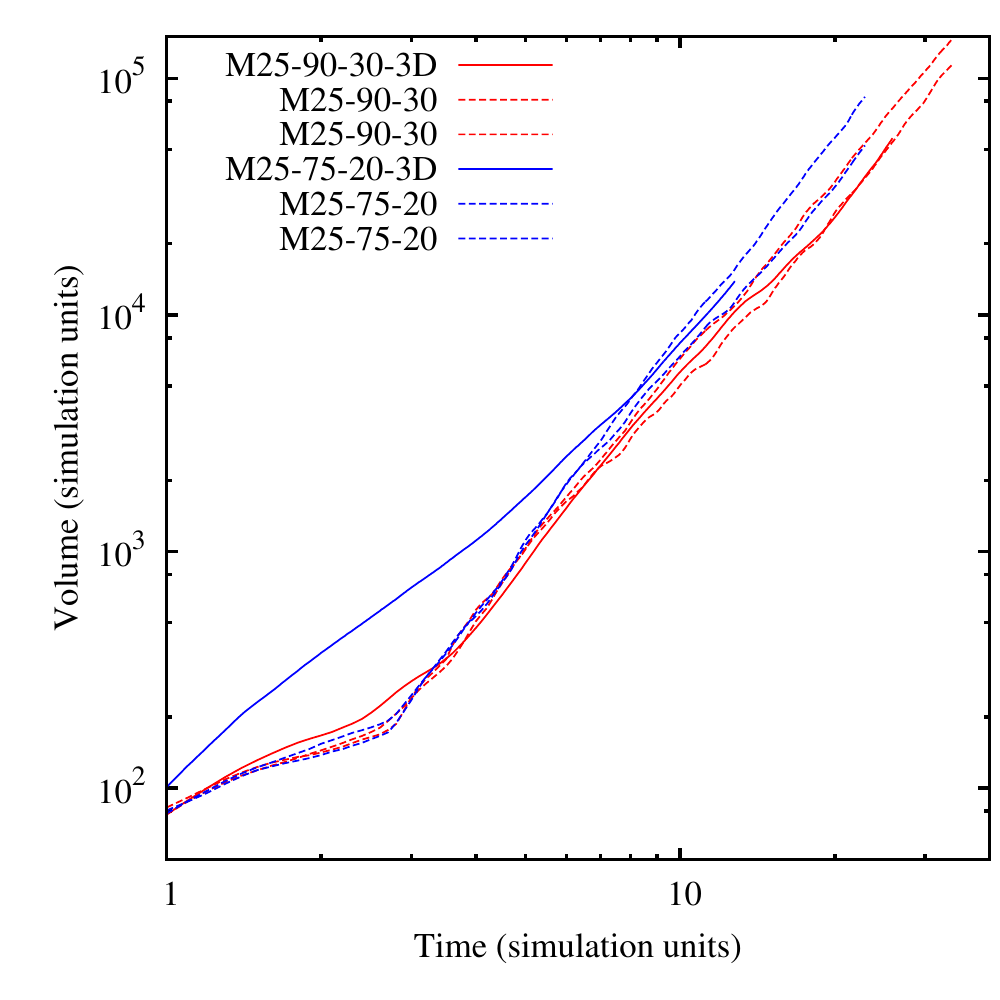}
\caption{Length and volume as a function of simulation time for
  the two 3D simulations (solid lines) and the two lobes of the
  corresponding 2D simulations (dashed lines).}
\label{lobe-3d}
\end{figure*}

\begin{figure*}
\includegraphics[width=17cm]{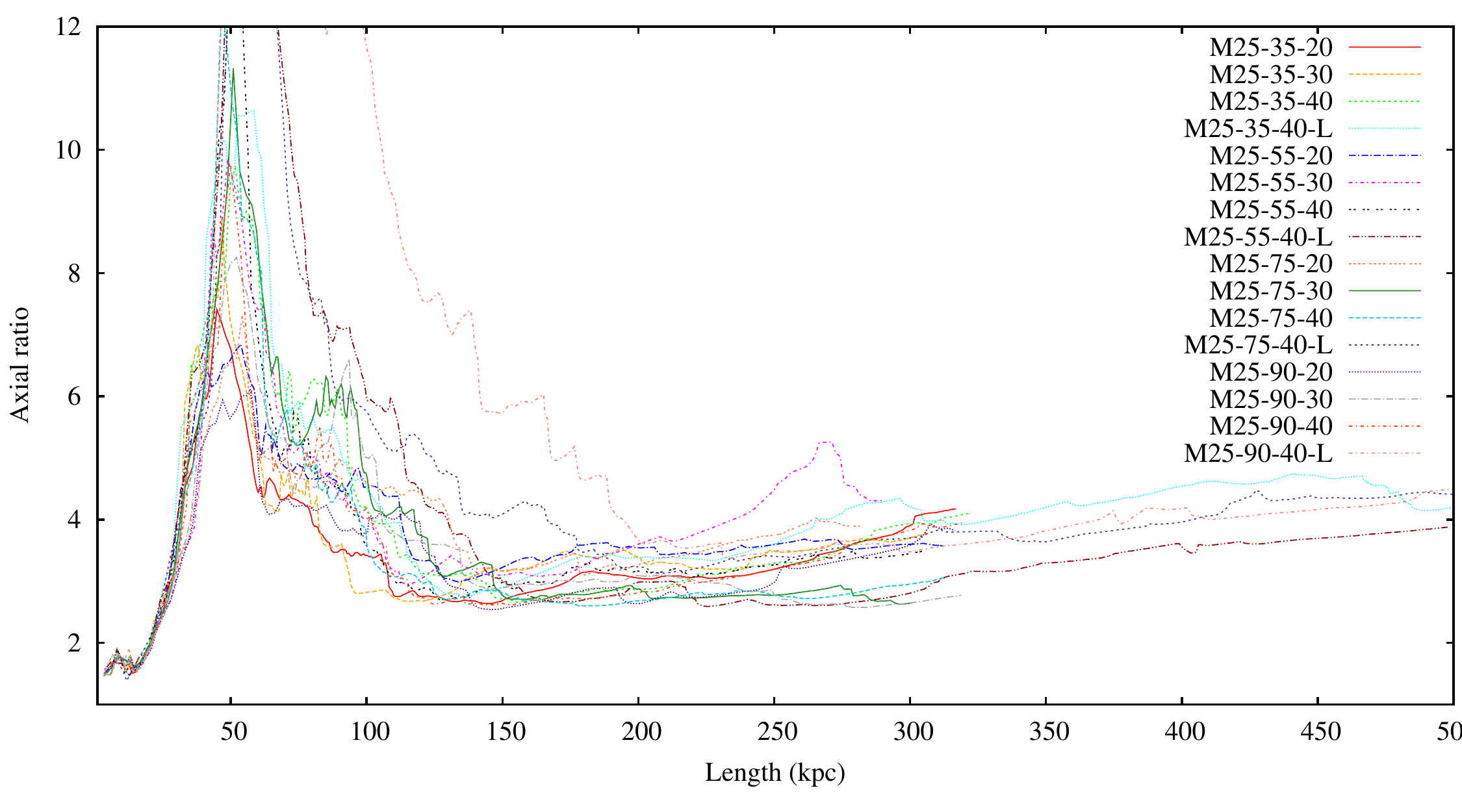}
\includegraphics[width=17cm]{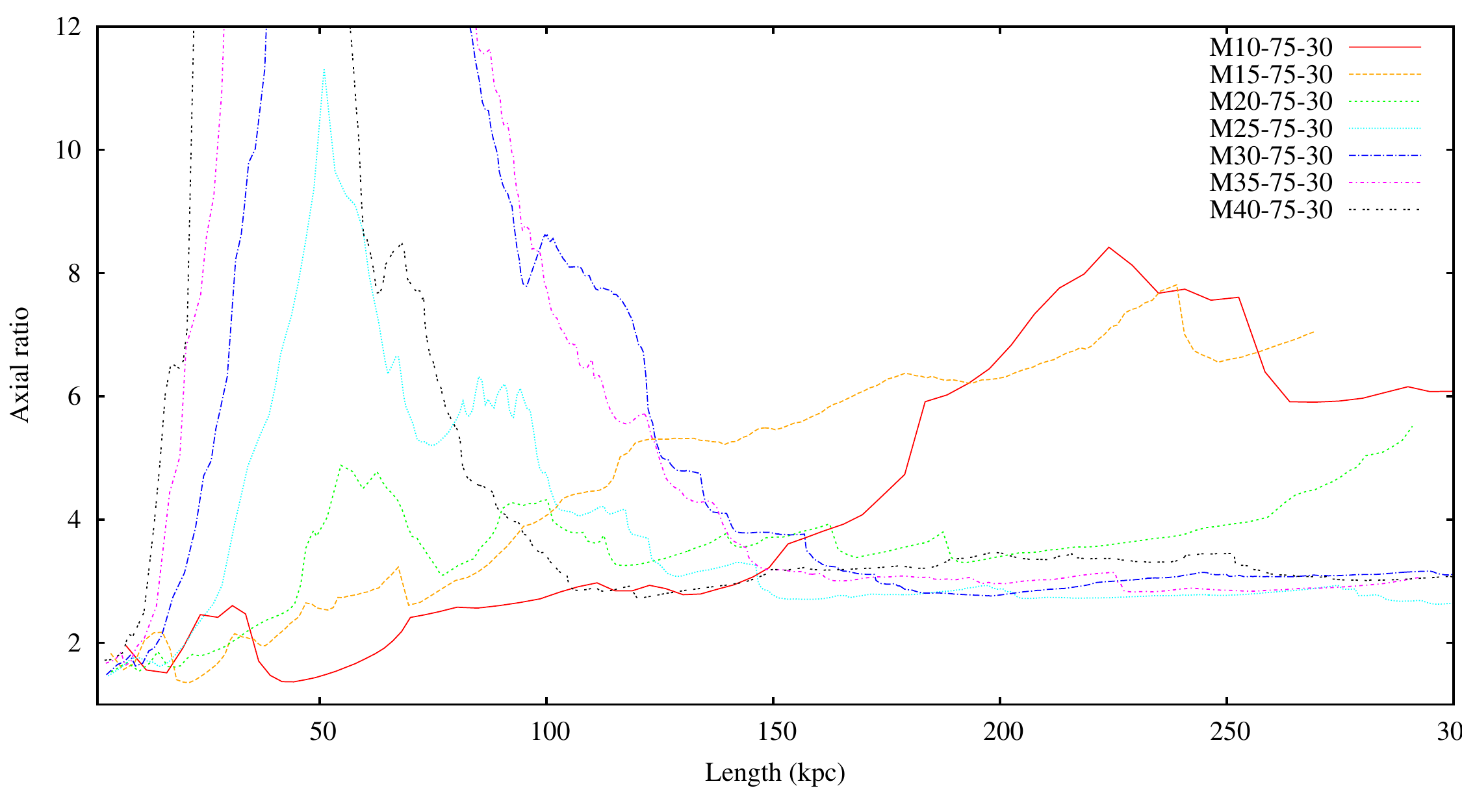}
\caption{Axial ratio as a function of lobe length in kpc for (top) all the
  ${\cal M}=25$ simulations and (bottom) simulations with varying Mach
  numbers but $\beta = 0.75$, $r_{\rm c} = 62$ kpc.}
\label{axial}
\end{figure*}

\begin{figure*}
\includegraphics[width=8.5cm]{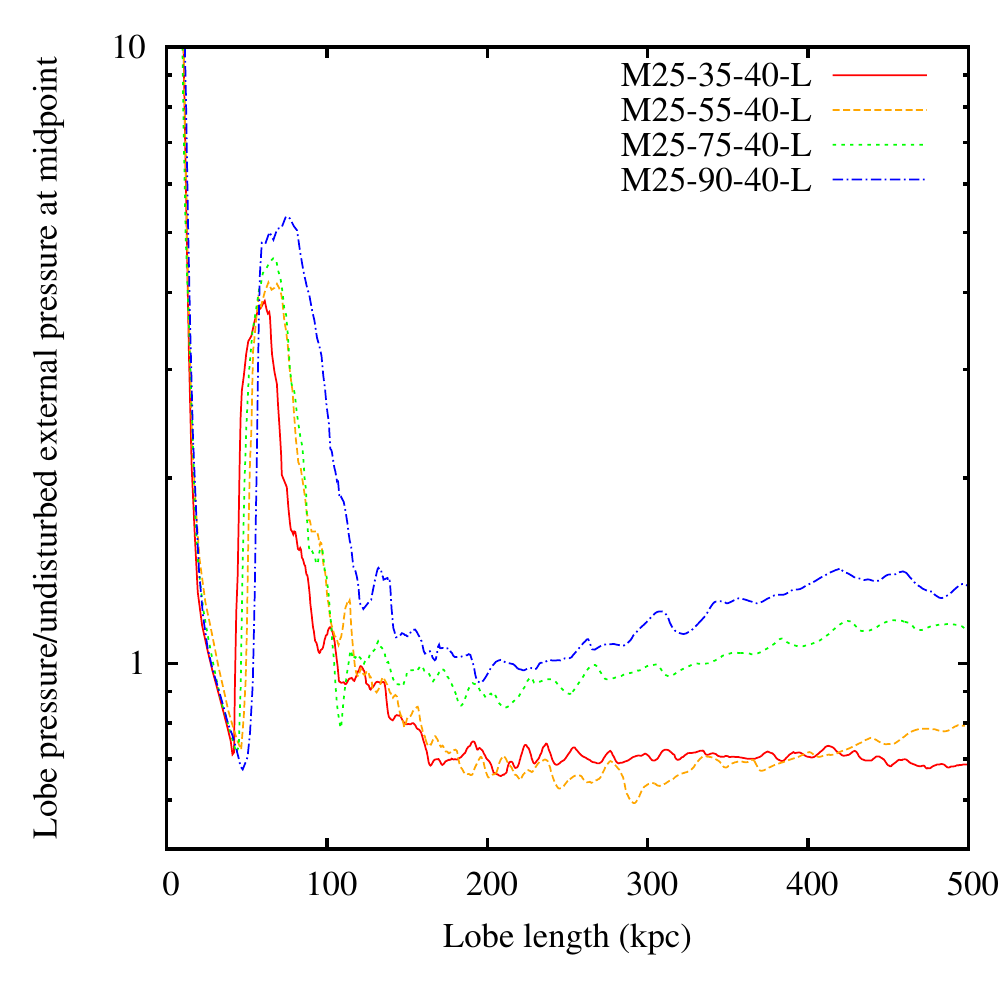}
\includegraphics[width=8.5cm]{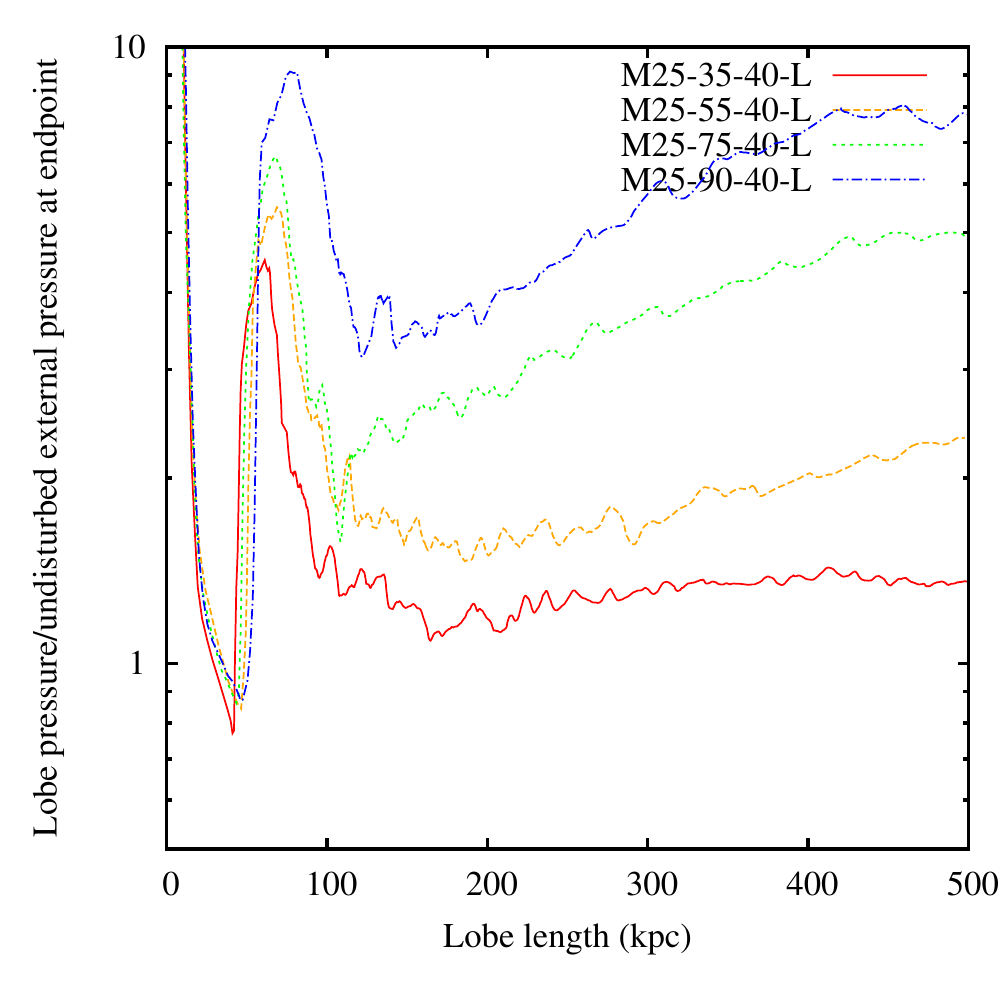}
\caption{The ratio between mean pressure in the lobes and {\it
    undisturbed} external pressure at (left) the lobe midpoint and
  (right) the end point of the lobes as a function of lobe length. Only
  the results from the long simulations are plotted here, but other
  runs are similar.}
\label{pressures}
\end{figure*}

In Figs \ref{length} and \ref{volume} we plot the mean length and
volume of the lobes in the ${\cal M}=25$ runs as a function of
simulation time. Here, for each simulation run, we have taken the
average length and volume of the two lobes in order to reduce the
scatter and highlight the interesting features. Considering the linear
growth of the lobes first (Fig.\ \ref{length}) we can generally pick
out three distinct slopes in the plots: an initial rapid growth (where
the jet grows quickly without forming much in the way of lobes, as we
noted above), followed by a phase of slower growth, followed by more
rapid expansion again. As the transition between these last two phases
happens on scales comparable to $r_{\rm c}$ (and in fact clearly
happens earlier for the sources with smaller $r_{\rm c}$; see
Fig.\ \ref{length}), it is natural to assume that it is the result of
the lobes emerging from the approximately uniform-density material
within $r_{\rm c}$ and starting to probe the approximately power-law
density regime seen on large scales. Consistent with this, we have
plotted the functional form of the length against time, $L_L \propto
t^{3/(5-3\beta)}$, from the analytic modelling of KA on these plots
(with arbitrary normalization) and we see that this is reasonably
comparable to the slope at late times -- this is particularly clear in
the four runs that go out to radii of 250 simulation units. (We note
that these four runs seem systematically slightly offset from the
equivalent runs with outer radii of 150 units -- we attribute this to
the slightly lower resolution of the larger runs, which seems to mean
that the lobe is slightly longer at all simulation times shown on
these plots.) Thus we reproduce not just the qualitatively expected
behaviour -- higher $\beta$ should mean faster growth at late times --
but also do surprisingly well at reproducing the quantitative
predictions of KA.

In the volume plots (Fig. \ref{volume}) we see a very similar
three-phase picture:
the lobe expands with relatively slow growth of volume at early times (we
explore the meaning of this below),
then grows rapidly after about $t=2$ in most simulations, and finally
expands more slowly and with roughly a power-law dependence on $t$.
The power law we see, though, is systematically flatter than the
prediction from KA, $V_L \propto L_L^3 \propto t^{9/(5-3\beta)}$. KA's
  prediction of course comes directly from their assumption of
  self-similarity in the lobes, so the fact that $V_L$ does {\it not}
  appear to scale as $L_L^3$ points to the fact, already seen in Section
  \ref{general}, that these lobes do not grow self-similarly.

The two 3D simulations show very similar lobe dynamics to their
  corresponding 2D simulations, as shown in Fig.\ \ref{lobe-3d}. Over
  most times the growth of length and volume in 3D is
  indistinguishable from the 2D results. The only noteworthy
  difference is that the volume seems systematically larger at very
  early times, particularly in the M25-75-20-3D simulation which has
  higher spatial resolution. This reinforces the impression that the
  initial rapid growth of the lobe is not well modelled, since the
  results depend on the numerical approach taken. However, the 3D and
  2D simulations agree very well after this initial phase.

The non-self-similar growth of the lobes is illustrated in another way
by Fig.\ \ref{axial} (top panel) which shows the axial ratios as a
function of lobe length for all the ${\cal M}=25$ sources in 2D
simulations. We define the axial ratio as the ratio of the lobe length
to the full lobe width, measured half-way along the lobe length: this
is very similar to the definition used by \cite*{Mullin+08}. With this
definition, large numbers imply thin lobes, small numbers correspond
to fat ones. Two important features are seen on this plot. Firstly, we
note the very strong peak in the axial ratios at small lobe lengths,
which corresponds to the phase where the source is growing rapidly in
length at close to constant lobe volume. This phase, which is probably
not realistic, is generally over by $L_L \sim 50L_1$ (or $\sim 100$
kpc) as the initial growth of the source slows down and the lobe
begins to inflate. More interestingly, at $L_L > 50L_1$ we see a
steady growth of the axial ratio as a function of lobe length; that
is, throughout most of the lobes' growth, the lobes are getting
thinner as they expand. This is because, while the growth in length of
the source is determined by the balance between the lobe pressure and
the momentum flux of the jet with the density and pressure at the end
of the lobes, the transverse expansion of the lobes is driven by the
internal lobe pressure only and the external density and pressure are
higher. The change in axial ratio with time or lobe length is not
consistent with self-similarity, but it is consistent with what is
observed \citep{Mullin+08}. Our simulations do not span the full range
of axial ratios observed in real sources, but they do lie in the right
region of parameter space, and of course a good deal of scatter is
introduced into observed axial ratios by projection effects.

Mach number has some effect on the axial ratio, in the sense that
simulations with very low ${\cal M}$ do not form realistic lobes at
any stage of the run; their lobe volumes remain low and so their axial
ratios are high. This is probably because the material in low
  Mach-number lobes is never shock-heated to high enough temperatures
  that the pressure in the lobes (which drives the transverse
  expansion) becomes comparable to the jet momentum flux. Changing the
  jet properties, e.g. the opening angle, or including some jet
  precession would probably allow lobes to form in some circumstances,
  but in any cases such low jet Mach numbers (for the jet powers we
  are considering) are not physically very plausible. For ${\cal M}
  \ge 25$, the late-time axial ratio appears more or less independent
  of Mach number, and is within the range seen in the various ${\cal
    M} = 25$ simulations (though we note that the early phase of fast
  lobe growth seems even more extreme for ${\cal M} > 25$).

\begin{figure}
\includegraphics[width=8.5cm]{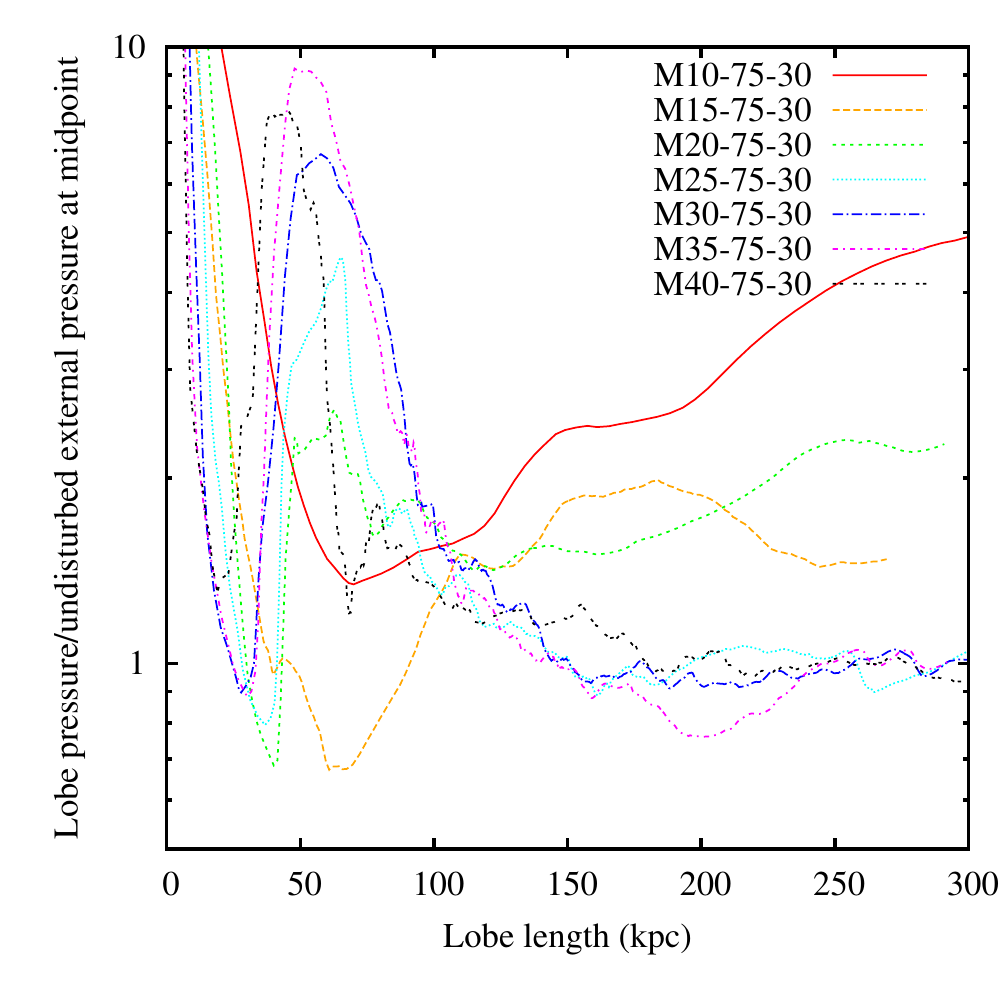}
\caption{The ratio between mean pressure in the lobes and {\it
    undisturbed} external pressure at the lobe midpoint as a function
  of lobe length in kpc for varying Mach number.}
\label{pressure-mach}
\end{figure}

Finally, we consider pressure balance in the lobes. We compare the
mean pressure in the lobes (since the sound speed in the lobes is
high, we do not expect any strong pressure gradients internal to the
lobes) to the {\it undisturbed} external pressure at the midpoints and
endpoints of the lobes. This approach is similar to what has been done
for radio galaxies, almost exclusively FRIIs, for which the mean lobe
pressure can be estimated from observations of inverse-Compton
emission and the external pressure from observations of thermal X-ray
emission \citep[e.g.][]{Hardcastle+02,Croston+04,Konar+09,Shelton+11}.
Observers use the undisturbed external pressure because this can
easily be estimated observationally, e.g. by masking out the lobes and
fitting spherically symmetric projected $\beta$ models, although of
course the lobes are not expected to be in direct contact with this
material. What we find is quite striking (Fig.\ \ref{pressures}):
after the initial expansion, and at $L_L \approx L_2$ the lobes come
into, and remain in, rough pressure balance (within better than a
factor 2) with the external pressure at the lobe midpoint, though they
are consistently strongly overpressured with respect to the external
pressure at the ends of the lobes (as expected since they are seen to
drive strong shocks at all times: Section \ref{general}). Thus the
simulations reproduce a key observational result. We also see (Fig.
\ref{pressure-mach}) that this result is independent of Mach number so
long as the Mach number is high enough that realistic lobes form
(${\cal M} \ge 25$). The 3D simulations reproduce these 2D results.

Overall, the simulations seem to do reasonably well at late times at
reproducing the behaviour we expect from real radio galaxies. At early
times, they are not realistic, as noted above. Only when the lobe has
grown to a size comparable to the core radius do we enter a regime
that is comparable to the behaviour of real radio galaxies. However,
once the simulations enter that regime, they seem to stay there,
independent of jet Mach number, so long as it is high enough, and of
the dimensionality of the simulation. This means that it is reasonable
to consider the energetic effects of the radio source in its
environment, as long as we confine ourselves to the period after $t
\sim 5$ or so when the lobes are fully established.

\subsection{Energetics of the lobe/environment interaction}
\label{energetics}

\begin{figure*}
\includegraphics[width=8.5cm]{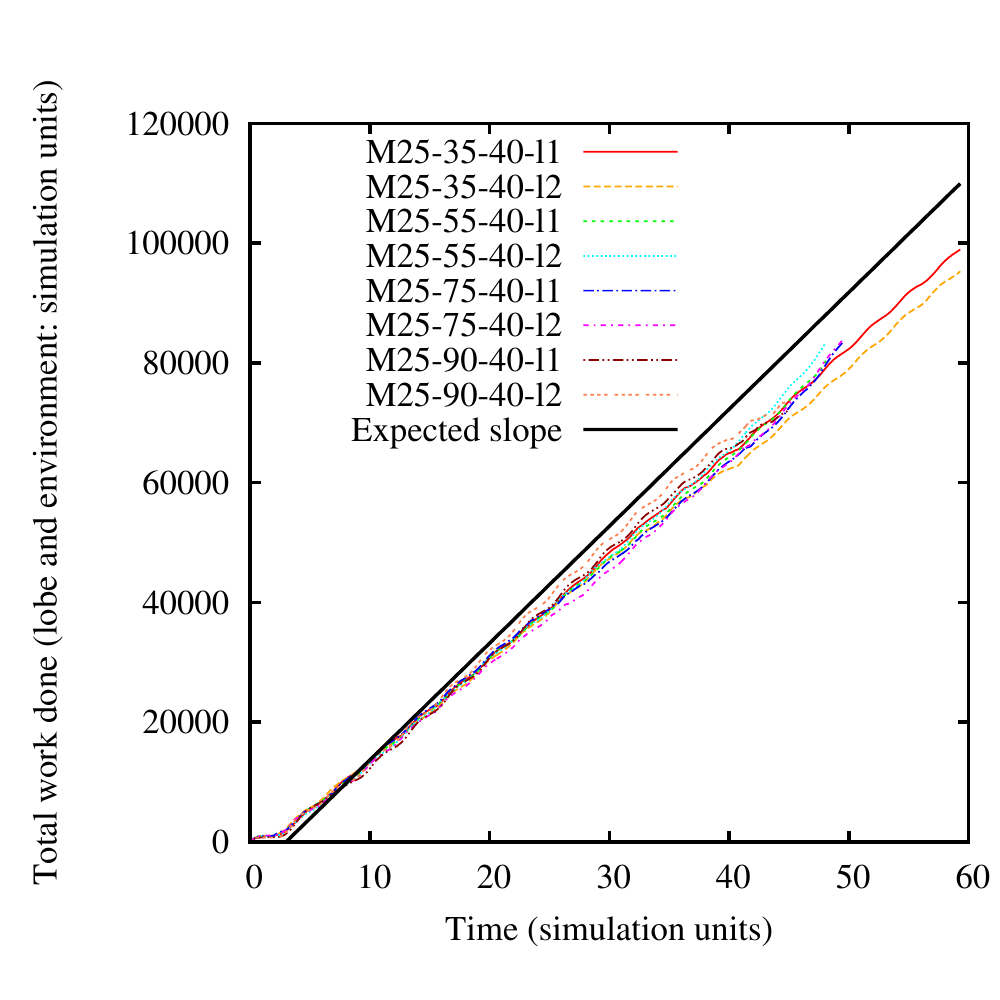}
\includegraphics[width=8.5cm]{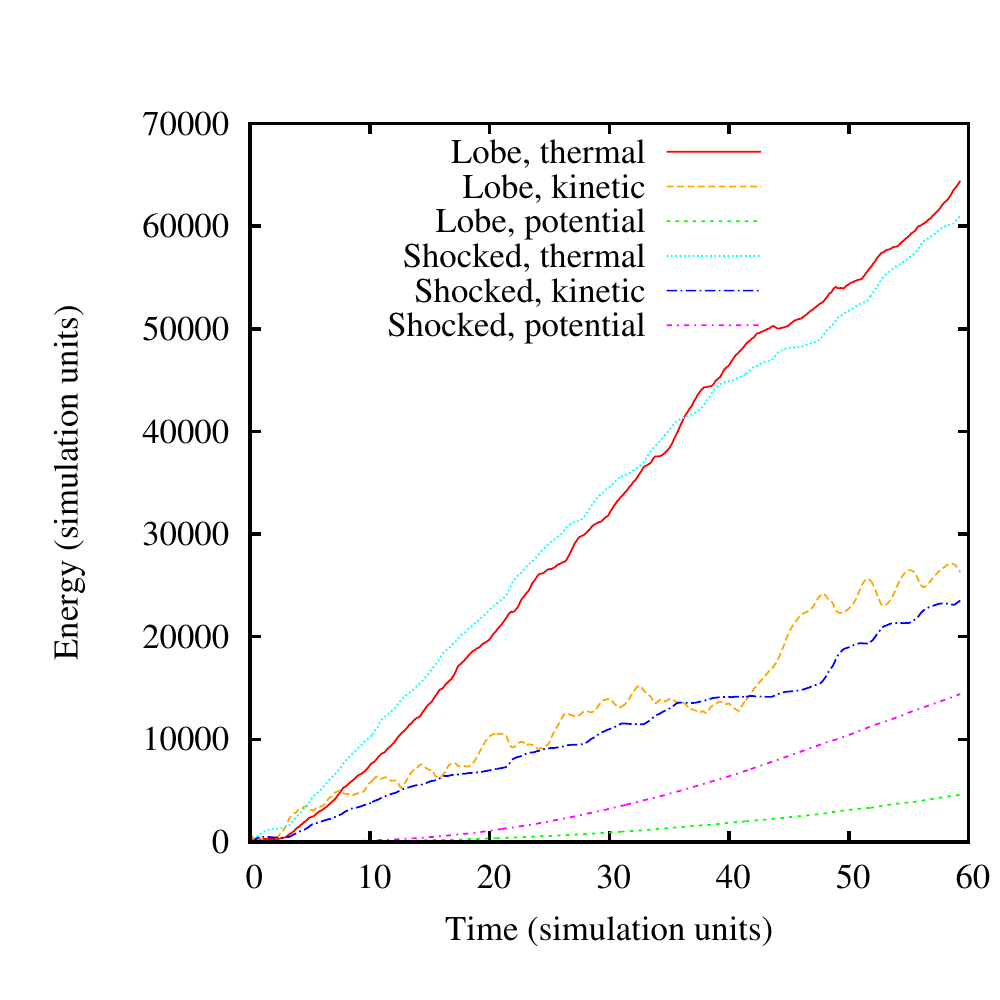}
\caption{Left panel: the total energy stored in the jets, lobes and external
  environment as a function of simulation time for each lobe of the
  four ${\cal M}=25$ simulations with $r_{\rm c} = 40$, together with the
  expected slope as discussed in the text. Right panel: the energy budget of the lobes and shocked region in the
  simulation M25-35-40 as a function of simulation time.}
\label{workdone}
\label{energies-example}
\end{figure*}

In this subsection we consider the energetics and environmental impact
of the lobes.

The first and most basic test is to check whether the jets are
carrying the expected level of energy into the simulation as a
function of time. The simulation energy unit $\epsilon$ is $\rho_0
L_1^3 c_s^2$, or around $4.6 \times 10^{49}$ J for our reference
simulations with ${\cal M} = 25$. So we can easily compare the rate of
growth of energy in the lobes and environment with the expected jet
power $Q_0$, or, equivalently, we can calculate the expected input
power in simulation units (${\cal M}^3/8$) to see whether the
simulations are conserving energy as expected. The results of such a
comparison are shown for some representative simulations in
Fig.\ \ref{workdone}. There are several points to note from this
figure. Firstly, we see that there is a short period at the start of
every simulation plotted where the total energy in the simulations
stays constant, rather than increasing. In detail, it seems that the
jet stops flowing into the simulation volume at these times: because
the jet is implemented as a boundary condition, it is possible for the
jet's entry on to the grid to be blocked, e.g. by backflow, with no
energetic consequences. Once the jet is established, though, we see an
approximately linear growth of energy with time, though note the
superposed oscillation which is a result of the slight sinusoidal
variation imposed on the Mach number (Section \ref{s-setup}); we plot
the two lobes separately in this figure so that the different phases
can be seen. Finally, we note that although the gradients in different
simulations are in close agreement, they disagree with the expected
slope at around the 10 per cent level, even after a non-zero start
time is imposed to account for the fact that a persistent jet only
starts at around $t=3$. As we see slightly different slopes for our
runs with slightly different resolutions, we believe this discrepancy
to be the result of resolution effects connected with the
implementation of the jet as a boundary condition. In particular, it
seems that the very edges of the jet have a lower density than they
should have even at the inner radius, so that less energy enters the
simulation than our simple calculation suggests. However, such
structures at the jet inlet should stay constant with time, and this
is consistent with the fact that the missing power is also constant
with time. If there were significant errors in conservation of energy
in the simulations, or in the calculations done in post-processing, we
would expect the curves in plots like those of Fig.\ \ref{workdone} to
be non-linear with time, given that the growth of the lobes and the
shocked regions is very clearly non-linear. As this is not the case,
we are confident that the simulations can be used to study the
energetic impact of the expanding lobes after early times; as noted
above, the dynamics of the lobe are not realistic at early times
anyway.

\begin{figure*}
\includegraphics[width=17cm]{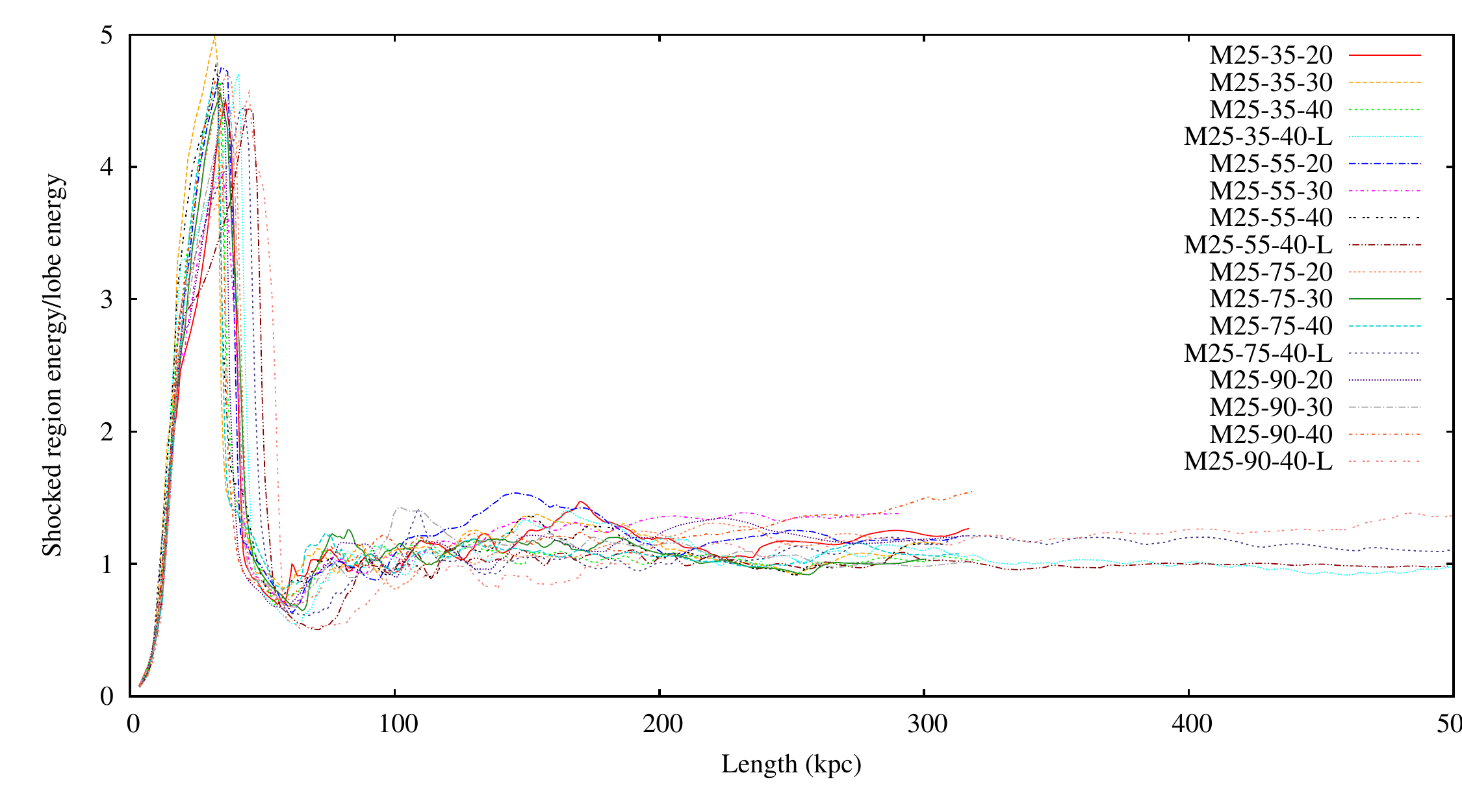}
\includegraphics[width=17cm]{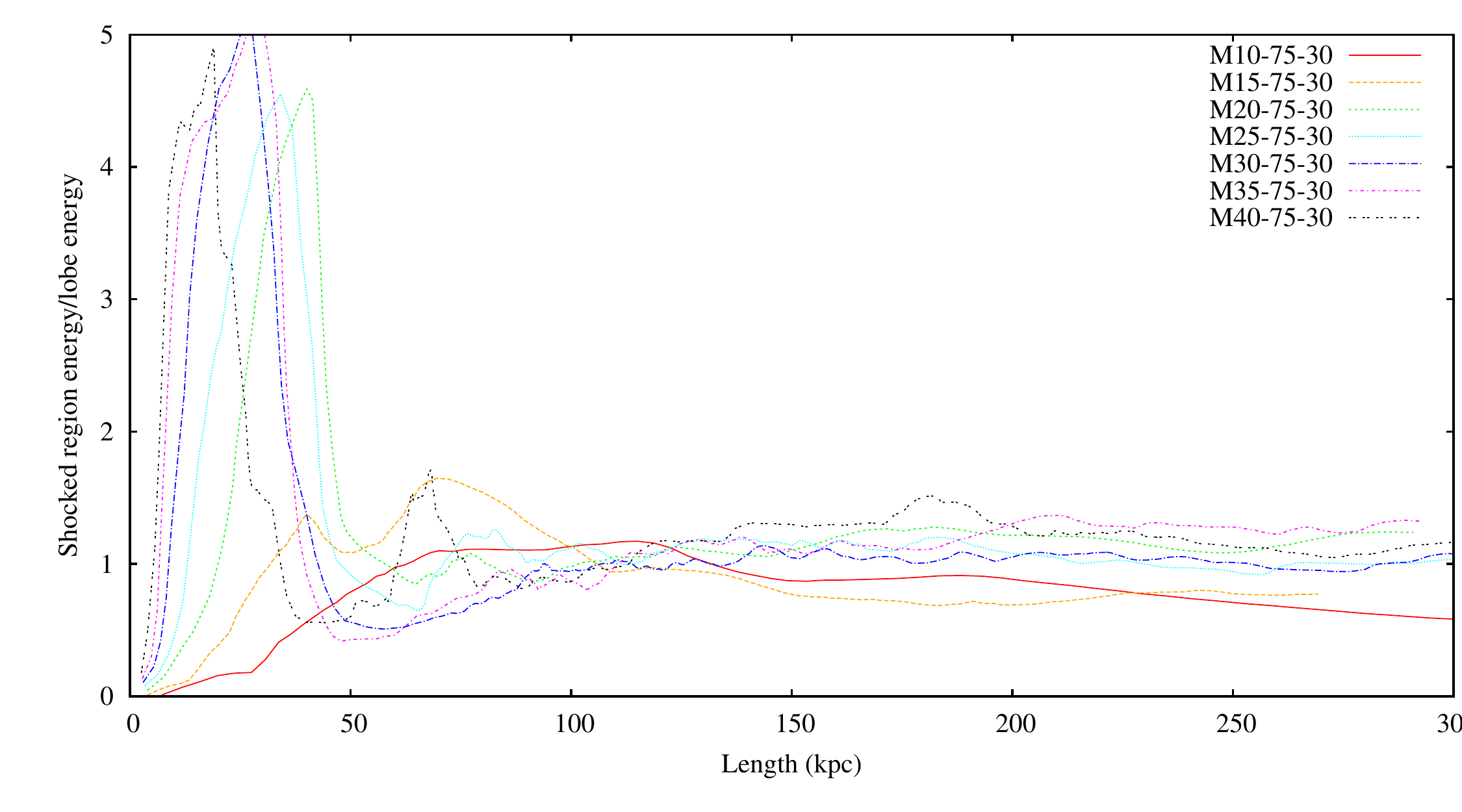}
\caption{The ratio between the energy stored in the shocked region and
the lobes as a function of lobe length; top, for all ${\cal M} = 25$
simulations; bottom, for simulations with different Mach number and
$\beta = 0.75$, $r_{\rm c}$ = 62 kpc.}
\label{eratio}
\end{figure*}

\begin{figure*}
\includegraphics[width=17cm]{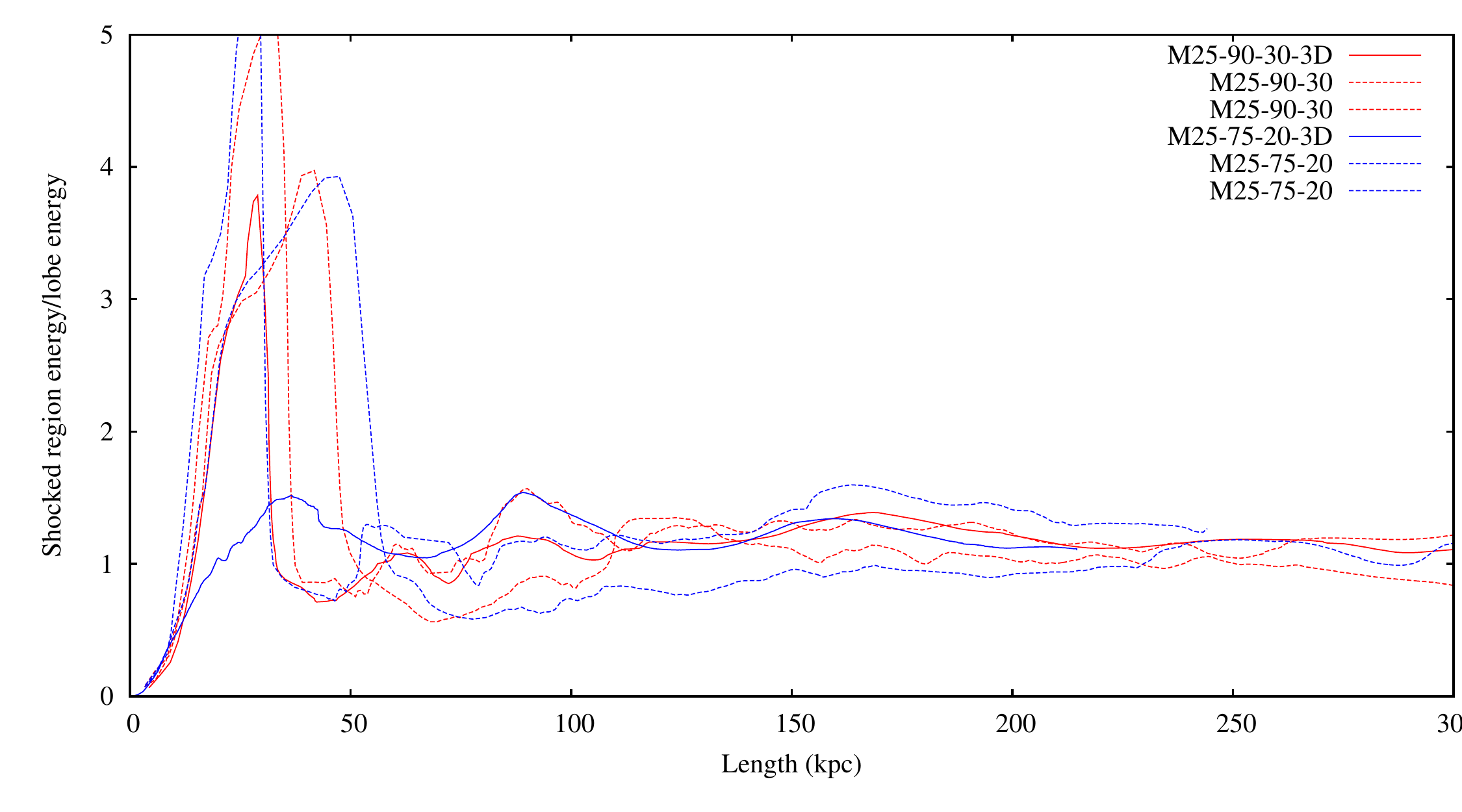}
\caption{As for the top panel of Fig.\ \ref{eratio}, but the two 3D
  simulations are shown compared to the results for individual lobes
  of the comparator 2D simulations. Solid lines show the 3D
  simulations and dashed lines the 2D ones. No significant differences between
the 2D and 3D results are seen after $L\sim 50$ kpc.}
\label{eratio-3d}
\end{figure*}

Our post-processing accounts separately for the thermal, kinetic and
potential energies in both lobes and in the corresponding shocked
regions. Thus we can investigate how the work done by the jet is
distributed between these different contributions to the total energy.
An example of this type of analysis is shown in
Fig.\ \ref{energies-example}. We see that, as might be expected, the
thermal energy in the lobes and the shocked region dominates over the
other components. Kinetic energy is non-negligible in both, however,
accounting for something like a third of the total. At late times the
potential energy of the shocked material, i.e. the work done in
lifting shocked material out of the centre of the galaxy, also becomes
important. There are strong variations in the kinetic energy of the
lobe material, which partly reflect the imposed variations in Mach
number of the injected jet, but also show the effects of the repeated
internal shocks in the lobes.

Perhaps the most obvious feature of this plot is that the thermal and
kinetic energy stored in the lobes keeps remarkably well in step with
that in the shocked region over most of the lifetime of the source. We
can quantify this for the simulated sources in general by plotting the
ratio of the total energy stored in the shocked region to that in the
lobes (Fig. \ref{eratio}). We see that at all lengths (times) after
the regular lobe growth is established ($L = 30L_1$, corresponding to
$t \sim 3$), the ratio of external to internal energies is close to
unity, with no very strong dependence on the parameters of the
environment (there does seem to be a weak dependence, in the sense
that this ratio tends to be higher for higher $\beta$ values). This
result confirms earlier findings in MHD simulations by
\cite{ONeill+05} and \cite{Gaibler+09}, but is based on a much more
comprehensive sampling of parameter space. We return to this very
important result in Section \ref{discussion}.

Like others we have considered above, this result appears more or less
independent of Mach number within the range we have used (Fig.
\ref{eratio}): even in the low-Mach-number systems where realistic
lobes are not formed, the energy ratio does not differ strongly from
unity at late times. It is also reproduced in the 3D simulations
(Fig.\ref{eratio-3d}).

\begin{figure*}
\includegraphics[width=17cm]{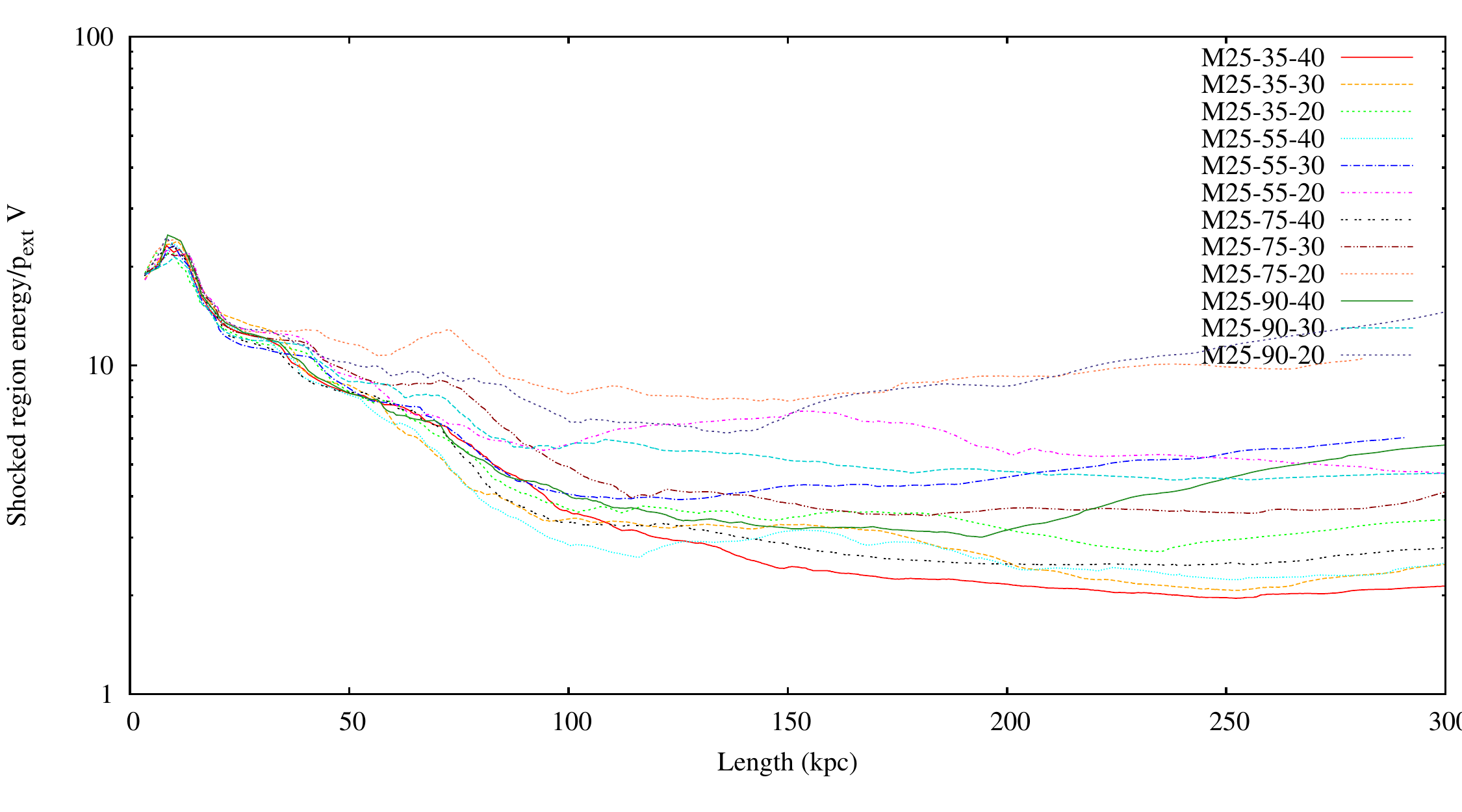}
\caption{The ratio between the energy stored in the shocked region and
$p_{\rm ext}V$ as a function of lobe length for the standard-sized ${\cal M} = 25$
simulations.}
\label{pvratio}
\end{figure*}

Finally, we can compare the work done on the external medium with
commonly used methods for estimating it. As we have seen, the internal
energy in the lobes tracks that in the shocked region reasonably well
over most of the lifetime of the source (Fig. \ref{energies-example}).
Therefore, if the internal energy of the lobe can be estimated, the
work done on the external medium can be estimated too. Many literature
estimates of the work done \citep[e.g.][]{Birzan+04} assume that
it is $\sim pV$, where $p$ is some relevant pressure and $V$ is the
volume of the lobe, or of the observed X-ray cavity. This must be of the
right order of magnitude, but setting aside the problems in estimating
$V$ in real radio galaxies, where projection effects are important,
the question is what value of $p$ to adopt. If the internal pressure
of the lobe (e.g. from inverse-Compton emission) is known, then this
is easy, but in this case of course we know the internal energy
density of the lobe as well. If the internal pressure is not known,
then our results above suggest that the undisturbed external pressure
at the lobe midpoint can give a reasonably good estimate of it. In
this case, we can use our simulations to see how badly in error a
simple $p_{\rm ext}V$ estimate is.

The ratio between the work done on
the external medium and $p_{\rm ext}V$ is plotted in
Fig.\ \ref{pvratio}. If the radio source were in pressure balance with
the external medium at the midpoint, and if all the energy in the lobe
were thermal energy, then we would expect this ratio to have a
constant value of 1.5 for roughly equal internal and external energy.
However, the fact that some of the work done on the lobes is in the
form of bulk k.e. and that the midpoint pressure balance is only
approximate causes deviations from this relationship; factors in the
range 2--10 are typical. In detail, the trends in Fig.\ \ref{pvratio}
are readily understood. At early times, all simulated radio sources
are strongly overpressured and drive strong shocks. Consequently, the
ratio is high; it decreases as the source comes close to pressure
balance. The simulations with smaller core radii and larger $\beta$
show higher ratios presumably because they are further from pressure
balance over most of their length; therefore, at late times, for our
smaller core radii and larger values of $\beta$ we generally find
values closer to 10, whereas for large core radii and small betas the
ratio tends towards two. To summarize, although the exact numerical
factor is likely to be different for real radio galaxies, with a
relativistic adiabatic index, it is safe to say that $p_{\rm ext}V$
pressure estimates are likely to underestimate the work done on the
external medium by a factor of a few; moreover, this factor will
depend on the age of the radio galaxy and the environment in which it
resides. Of course, it should be noted that our results apply
  principally to FRII radio galaxies, while many of the objects
  showing cavities are lower-power FRIs; although qualitatively we
  might expect similar caveats to apply to a $p_{\rm ext}V$ analysis
  of FRIs, our simulations do not allow us to make quantitative
  statements about such systems.

\subsection{Radio emission}
\label{radio}

\begin{figure*}
\includegraphics[width=17cm]{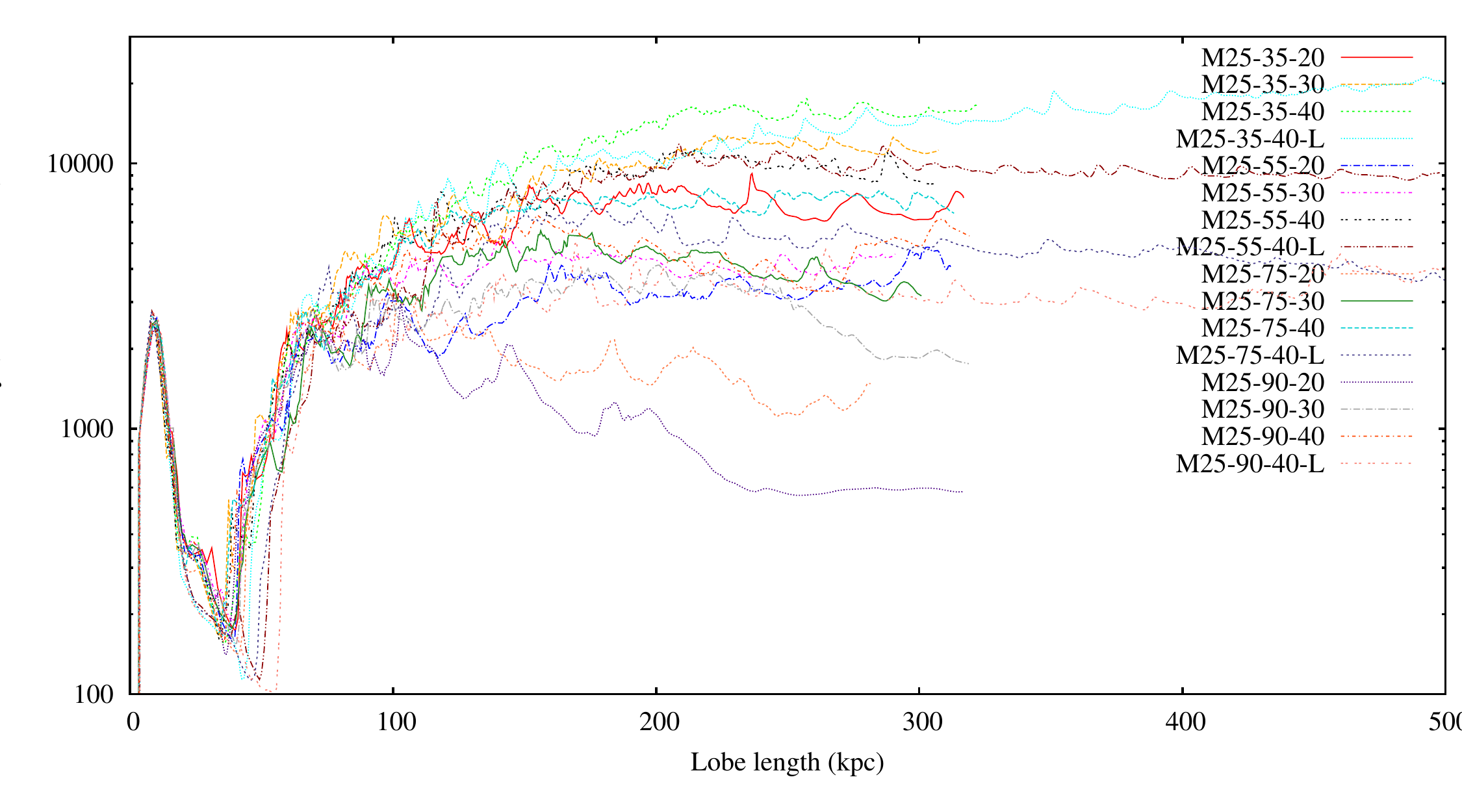}
\caption{Radio luminosity (simulation units) as a function of mean
  lobe length for the ${\cal M}=25$ simulations.}
\label{radio-sim}
\end{figure*}

Our simulations do not contain all the physics needed (i.e., electron
acceleration, loss processes and magnetic fields) to carry out a
proper visualization of the synchrotron emission from the lobes. It is,
however, interesting to ask how the overall radio luminosity of the
lobes would evolve with time, and we can do this by assuming that
the pressure that we measure in the lobes is contributed by electrons
and magnetic field. Since the simulations do not constrain the
contributions of these separately, we need to assume equipartition or
some constant departure from equipartition (as measured by
inverse-Compton observations, e.g. \citealt{Croston+05}). Let the
energy density in electrons be $U_e$ and that in magnetic field be $U_B$, then this assumption corresponds
to
\[
\eta U_e = U_B = \frac{B^2}{2\mu_0}
\]
where $\eta$ describes the energetic departure from equipartition,
$\eta < 1$.

\begin{figure*}
\includegraphics[width=17cm]{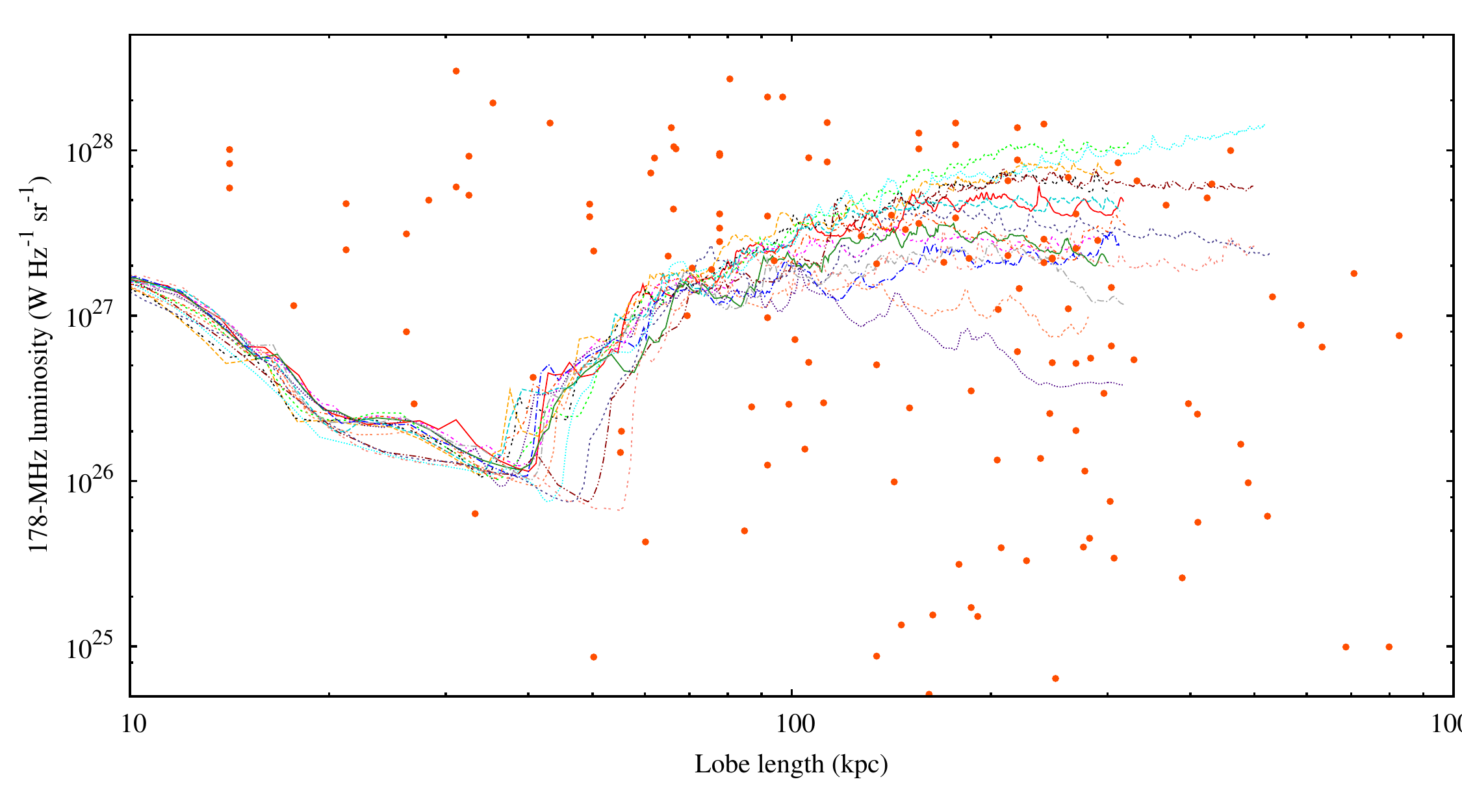}
\caption{Radio luminosity (physical units) as a function of mean lobe
  length for the ${\cal M}=25$ simulations, converted from simulation
  units using the factor ${\cal L}$ for $\eta = 0.1$ as described in
  the text. The line styles correspond to the same simulations as in
  Fig.\ \ref{radio-sim}. Overplotted are the 3CRR radio galaxies above
  the nominal FRI/FRII cutoff of $5 \times 10^{24}$ W Hz$^{-1}$
  sr$^{-1}$, where we have scaled down the largest angular size by a
  factor $\sqrt{2}$ to account for the fact that the simulations are
  for a single mean lobe length and that a typical 3CRR source will be
  projected. Note that the different curves represent radio sources
  with the {\it same} jet power, but in different environments. The
  radio luminosity increases in environments with a larger core radius
  and shallower density decline.}
\label{radio-phys}
\end{figure*}

If we then assume a power-law electron distribution, it can be shown
using standard textbook results \citep[e.g.][]{Longair10} that the
volume emissivity goes as $p^{(q+5)/4}$, where $q$ is the electron
energy power-law index; the luminosity can be calculated by
integrating this over the lobe volume. Taking $q=2.2$, corresponding
to a low-frequency spectral index $\alpha = 0.6$, the emissivity
scales as $p^{1.8}$. We begin therefore by plotting $\int p^{1.8} {\rm
  d}V$ over the simulation volume as a function of lobe length
(Fig.\ \ref{radio-sim}). These plots correspond to the `tracks in the
$P$--$D$ diagram' commonly used in studies of the evolution of radio
galaxies, albeit still in simulation units at this point. We see first
of all that the `luminosity' in these plots varies strongly with
source length (and therefore time). Ignoring the strong variation
before the lobe is properly established at $L<50$ kpc, we see a
general trend for the luminosities first to scale up and then to level
out. At the longest lengths, the luminosities may level out or
decrease slightly, the most obvious decreases being seen in
environments with high $\beta$ and small $r_{\rm c}$. At $L\sim 200$
kpc, there is almost an order of magnitude difference in the radio
luminosities of sources with, by construction, identical jet powers.
We also note the smaller-scale, random or slightly periodic variations
in the radio power; these reflect fluctuations in the pressure in the
lobes and should not be taken too seriously, since the detailed
hydrodynamics in the lobes is probably not correct.

It is important to note that the analysis described here takes
no account of radiative losses. The approximation that the electron
energy distribution is a power law is not a bad one even when
radiative losses are effective, because the energy density in the
electrons is dominated by low-energy electrons with long loss
timescales; but of course the frequency $\nu$ needs to be low in order
to sample the part of the synchrotron spectrum that is still
adequately represented by a power law. Since we wish to compare
these radio luminosities to real luminosities, we use a notional
observing frequency of 178 MHz, which should be low enough to
  avoid significant spectral ageing effects in all but the oldest
  sources, and allows direct comparison to
observations of the 3CRR sample and to the work of \cite*{Kaiser+97}.

Given that the simulation unit of pressure is $P = \gamma n_0
kT$, or $1.6 \times 10^{-11}$ Pa, and the unit of volume is $L_1^3$,
it can be shown that the simulation unit of radio luminosity in
physical units (W Hz$^{-1}$ sr$^{-1}$) is
\[
{\cal L} = c(q) {\frac{e^3}{\epsilon_0 c m_e}} \left({\frac{\nu m_e^3 c^4}{e}}\right)^{-
\frac{q-1}{2}} \frac{3P^{\frac{q+5}{4}}}{4\pi I}  (6\mu_0\eta)^{\frac{q+1}{4}}
(1+\eta)^{-\frac{q+5}{4}} L_1^3
\]
where $e$ is the charge on the electron, $m_e$ its mass, $c$ the speed
of light, $\epsilon_0$ the permittivity of free space, $\mu_0$ the
permeability of free space, $I$ the integral of $E N(E)$ over the
energy range of the electrons, $E_{\rm min}$ to $E_{\rm max}$, and
$c(q)$ a dimensionless constant, the product of a number of gamma
functions, of order 0.05. Taking $q=2.2$, $\nu = 178$ MHz, $\eta =
0.1$ (given that the measured magnetic field strengths can be up to a
factor of a few below equipartition, see \cite{Croston+05}, $E_{\rm
  min} = 10m_ec^2$, and $E_{\rm max} = 10^5m_ec^2$ (the results are
insensitive to this choice of electron energy range), we find that
${\cal L} = 6.6 \times 10^{23}$ W Hz$^{-1}$ sr$^{-1}$ for the ${\cal
  M} = 25$ simulations. If we take $\eta = 1$, the conversion factor
roughly doubles, ${\cal L} = 1.4 \times 10^{24}$ W Hz$^{-1}$
sr$^{-1}$, so that we are not excessively sensitive to equipartition
assumptions.

This conversion factor allows us to generate a plot of the evolution
of radio luminosity in physical units for the ${\cal M} = 25$ sources
(results for other Mach numbers are similar but are omitted for
clarity). We overplot data for the 3CRR
sources\footnote{Data from http://3crr.extragalactic.info/ .}
(Fig.\ \ref{radio-phys}; note the change of $x$-axis scale from linear
to logarithmic). We see immediately that the luminosities are in the
right general area of parameter space for powerful radio galaxies; of
course, the 3CRR FRIIs we have overplotted should represent a wide range
of jet powers and environments, down to objects with $Q \sim 10^{36}$
W in very poor groups \citep[e.g.][]{Hardcastle+12}, so it is not
surprising that the simulations do not sample the full $P-D$ diagram.
Within the limitations of the simulations and the calculation we have
carried out (i.e., a comparison is only possible for $L > 50$ kpc, and
we neglect losses, which may well be significant on scales $>300$ kpc)
we view the position of our simulation tracks on this plot as good
evidence that we are generating physically reasonable radio sources.
We note that the typical luminosity, $\sim 10^{27.5}$ W Hz$^{-1}$
sr$^{-1}$, is exactly what we would expect from a $10^{38}$ W jet
using the results of \cite{Willott+99}, which, if a coincidence,
is certainly a remarkable one.

Having said this, it is striking that the most luminous sources in the
simulations (those located in large-$r_{\rm c}$, low-$\beta$
environments) approach the highest luminosities seen for 3CRR sources,
which is perhaps surprising given that we are only trying to simulate
intermediate-power FRIIs with $Q = 10^{38}$ W; however, these
environments are very rich, perhaps unrealistically rich for most
low-power 3CRR sources, and so we do not regard this as a significant
problem. We also note that any non-radiating particle content in the
lobe (see \citealt{Hardcastle+Croston10} for a discussion of the evidence
for these in FRIIs) would tend to reduce the required electron number
densities and hence the radio luminosity.

Finally, it should be noted that the very strong dependence of
luminosity on environmental properties on 100-kpc scales, which is
independent of these caveats, provides both a quantitative estimate of
the boosting effect of rich environments discussed by \cite{Barthel+Arnaud96} and a useful warning of the dangers of attempting to
infer jet power from radio observations alone. On these scales, there
is at least an order of magnitude scatter in the radio luminosity
expected for a fixed jet power. We return to this point in Section
\ref{discussion}.

\subsection{X-ray emission}
\label{xray}

\begin{figure*}
\includegraphics[width=17cm]{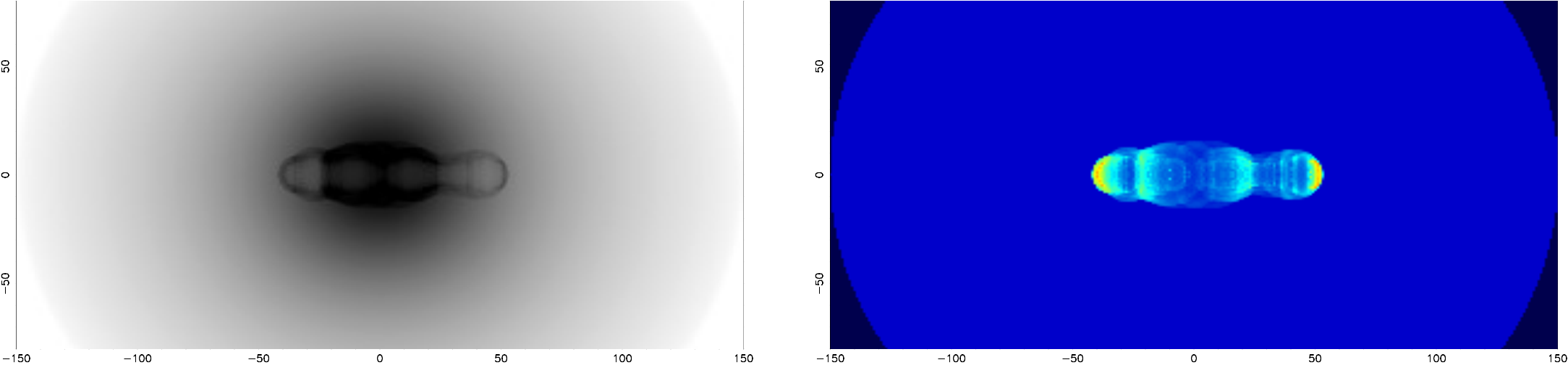}
\includegraphics[width=17cm]{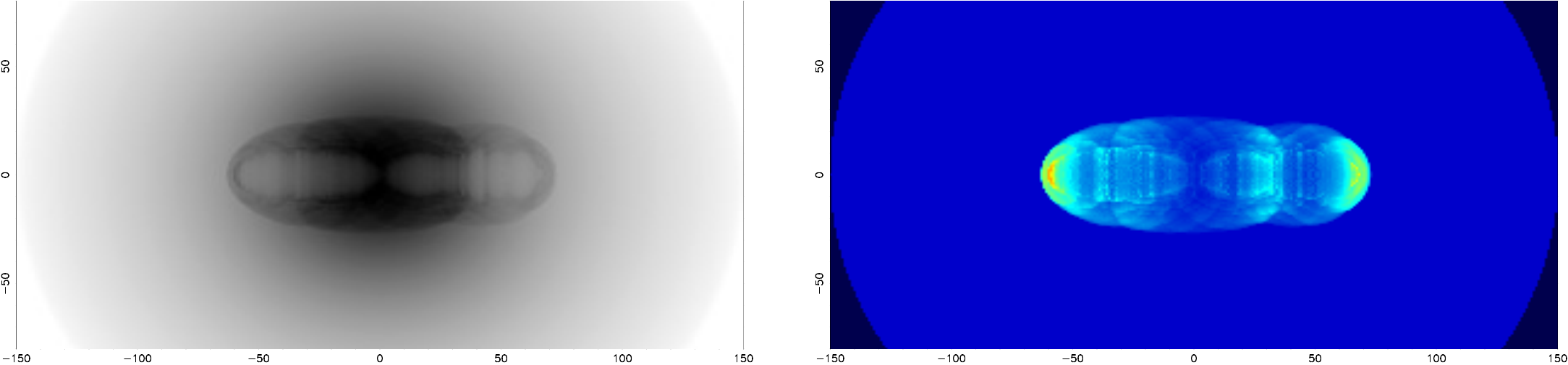}
\includegraphics[width=17cm]{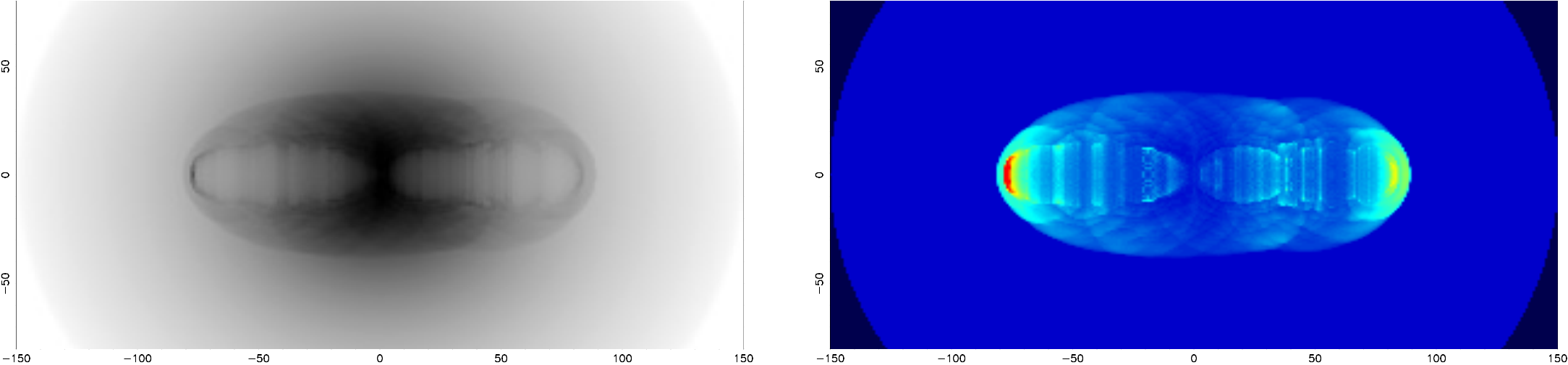}
\includegraphics[width=17cm]{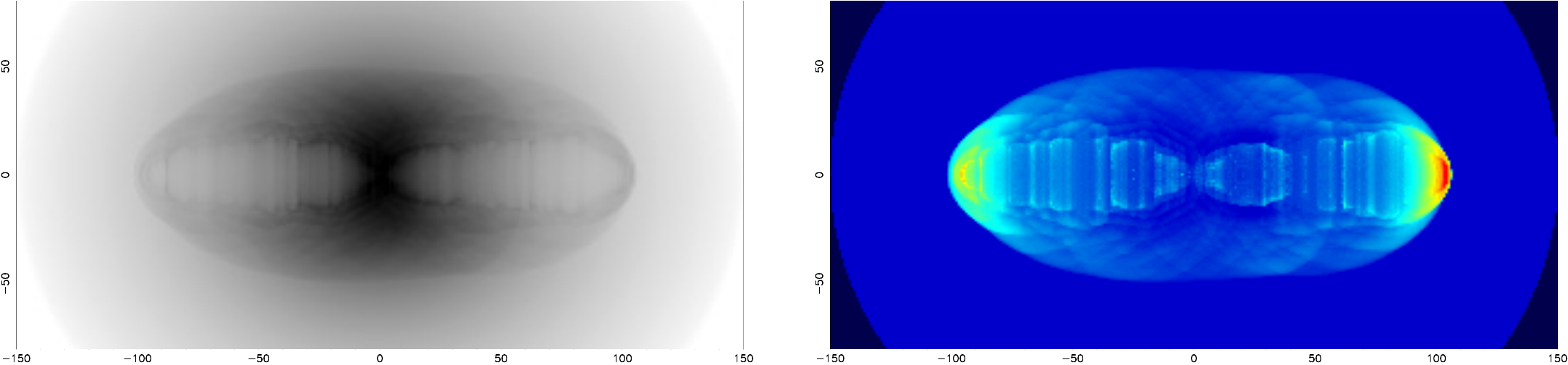}
\includegraphics[width=17cm]{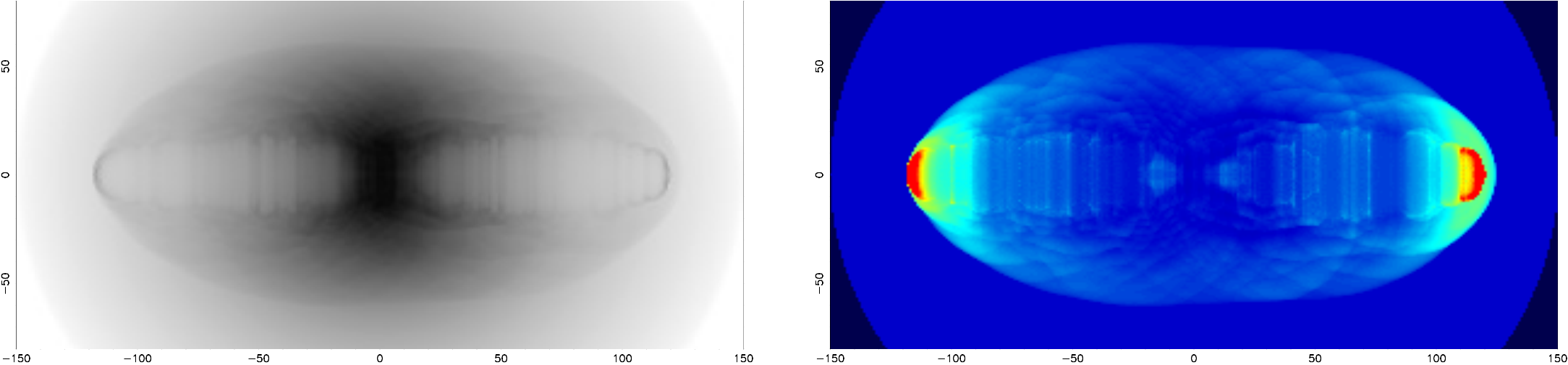}
\caption{X-ray visualization of the simulation M25-35-40 for
  simulation times (from top to bottom) $t=10$, 20, 30, 40 and 50. The left-hand images
  show {\it Chandra} counts in the energy range 0.5--5.0 keV, with a
  linear transfer function. The
  right-hand images show the corresponding emission-weighted
  temperature, with dark blue being coldest (1.8 keV) and red being
  hottest ($>3$ keV); see the text for discussion. The jet axis is at
  90$^\circ$ to the line of sight and the image resolution is 1
  simulation unit ($\sim 2$ kpc). Compare Figs \ref{density-example}
  and \ref{temperature-example}, which show slices through the
  midplane of this simulation at high resolution. Co-ordinates shown
  are simulation units.}
\label{xray-vis-90}
\end{figure*}

\begin{figure*}
\includegraphics[width=17cm]{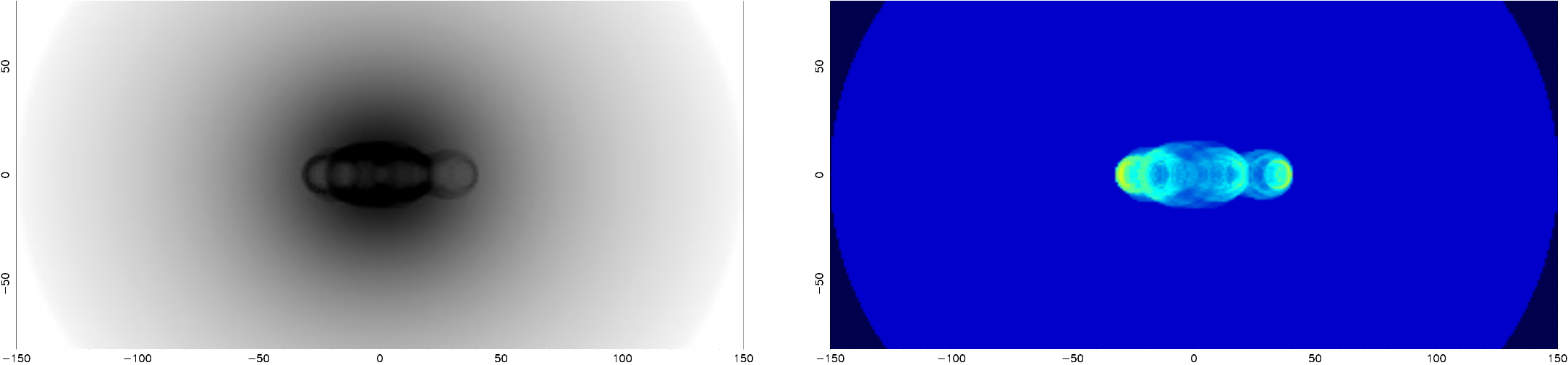}
\includegraphics[width=17cm]{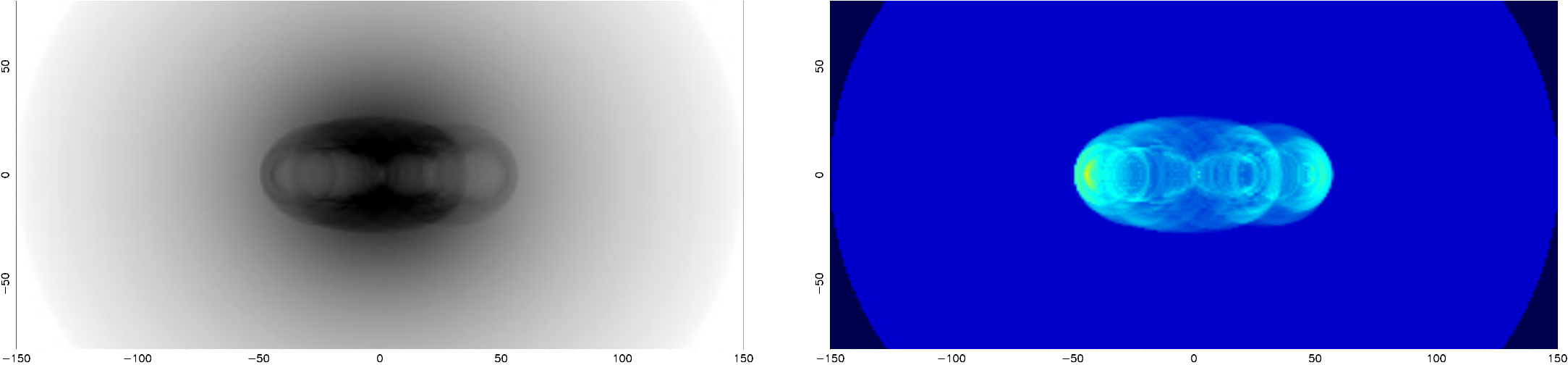}
\includegraphics[width=17cm]{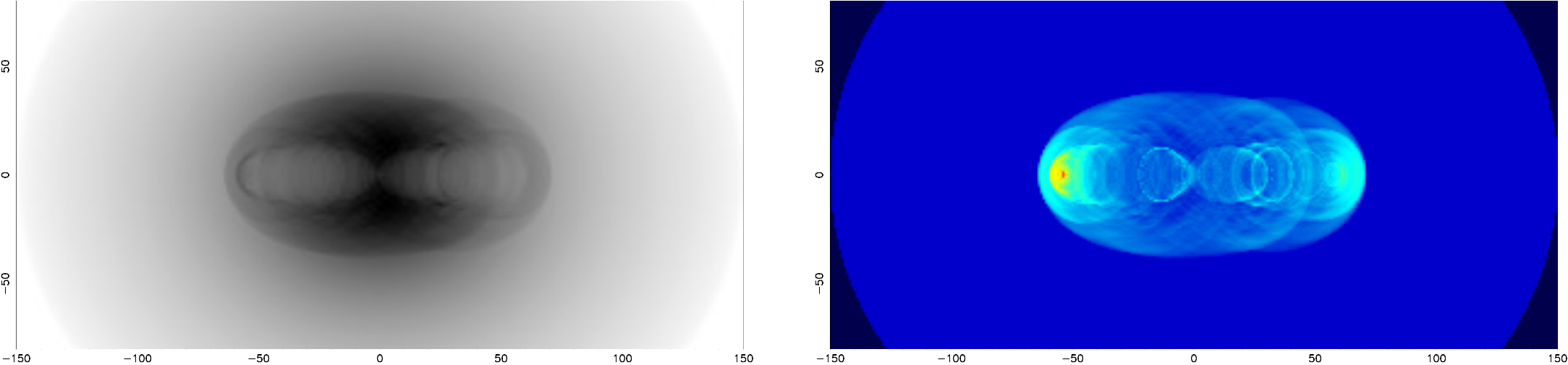}
\includegraphics[width=17cm]{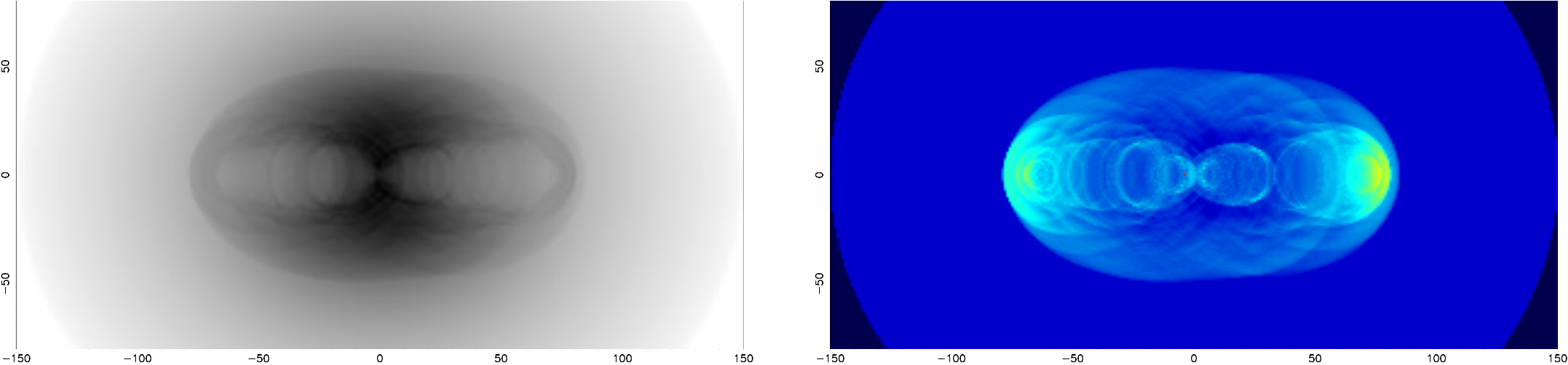}
\includegraphics[width=17cm]{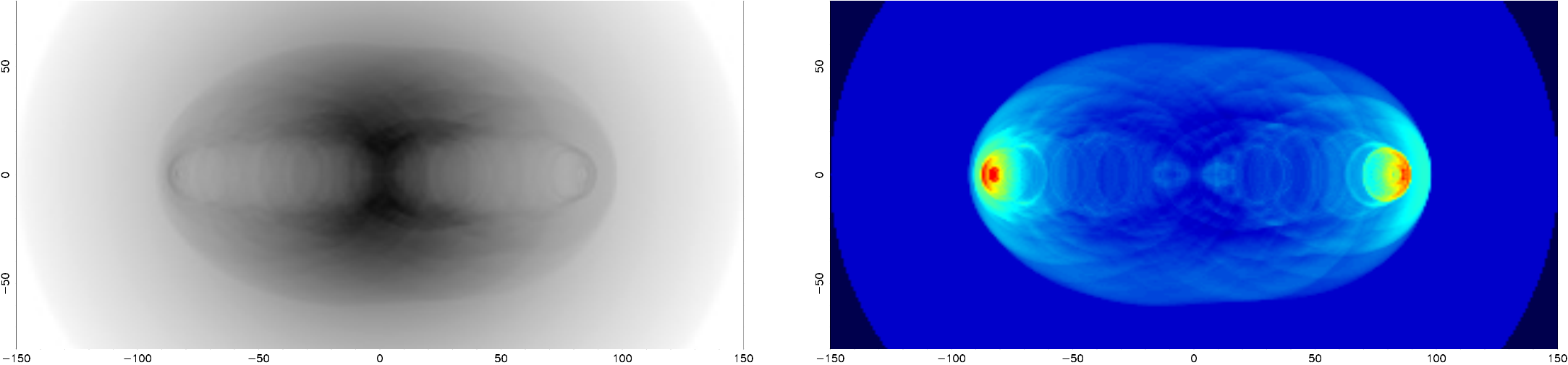}
\caption{As for Fig.\ \ref{xray-vis-90}, but the jet axis is at
  45$^\circ$ to the line of sight.}
\label{xray-vis-45}
\end{figure*}

We can assess the observable effect of the radio galaxy's interaction
with its environment in these models by visualizing the X-ray emission
from the environment. We choose to model X-ray emission as seen by
{\it Chandra}: {\it Chandra}'s angular resolution of $\sim 0.5$ arcsec
gives a spatial resolution of $\sim 2$ kpc at around $z=0.2$, so we
adopt a spatial resolution of 1 simulation unit for the ${\cal M} =
25$ simulations. We assume that lobe material does not contribute to
the X-ray emission, and therefore exclude it using the tracer. We use
{\sc xspec} with a real {\it Chandra} response file to model the
conversion between emission measure and (on-axis) count rate in the
0.5--5.0 keV energy band as a function of temperature for an APEC
model with 0.3 solar abundance (neglecting Galactic absorption). We
can then compute the expected X-ray image (for infinite
signal-to-noise) by integrating along lines of sight through the
simulated volume: we neglect emission outside the simulation
(corresponding to perfect background subtraction). We can also compute
the emission-weighted (strictly, counts-weighted) mean temperature
along the line of sight; {\it Chandra} does not quite measure this
quantity, but it is close enough to give an idea of the expected
observational effects. We show representations of the counts image and
the map of emission-weighted temperature for several times during the
simulation M25-35-40 and for two different angles to the line of sight
in Figs.\ \ref{xray-vis-90} and \ref{xray-vis-45}.

Bearing in mind that we are `observing' with infinite signal to noise
and that therefore many of the details seen in the simulations are not
likely to be visible in reality, these visualizations are
qualitatively very consistent with real observations. We note first of
all that, though there is an approximately elliptical shock with a
clear surface brightness contrast at the boundary, the measured
temperature difference across the shock is not likely to be very
great; the emission-weighted temperature of the shocked region is
$\sim 2.5$ keV compared to the unshocked 2-keV gas. This compares very
well to the modest temperature increases seen in this region in Cygnus
A \citep{Wilson+06}. The regions of strong shocks and high
temperatures ($>3$ keV) are confined to the very tips of the lobes,
and even then are not always present; it is easy to imagine that these
high temperatures will be hard to observe in real radio galaxies,
where measurements of spectra require integration over large regions.
One is more likely to see moderately increased temperatures over a
larger region at the head of the lobes, as suggested in the case of
Cygnus A by \cite{Belsole+Fabian07} and in 3C\,452 by \cite{Shelton+11}. Cavities are visible, but again quite low in surface
brightness contrast, and so not necessarily expected in real
observations (particularly as we have not attempted to model
inverse-Compton emission from the lobes, which in real observations
will fill in some of the cavity emission). The most obvious feature of
the simulated observations, which should be present in any real
observation if these simulations are well matched to reality, is the
emergence of a bright central `bar' of X-ray emission, together with
`arms' extending along the inner part of the lobes. This material is
cool, with a temperature equal to, or even in places slightly less
than, the original temperature; it is generated from the pre-existing
dense core of the host cluster, which is initially heated by the
radio-lobe shock but then cools adiabatically as the shocked region
expands ahead of it, in the manner discussed by \cite{Alexander02}.
(Radiative cooling of the thermal material might be expected to
enhance the density contrast here still further.) Structures like
this, positioned roughly inside the boundary of the radio lobes
(though note that projection will affect the lobes and the bar/arms
structure differently, so that we do not expect an exact match between
the two) are a smoking gun for a system in which the lobes have come
into pressure balance at large radii and are now being squeezed out as
in Model C of \cite{Scheuer74}. In real observations, structures that
could be interpreted in this way are in fact ubiquitous in
observations of FRIIs; examples include the central, bright, cool
structure in Cygnus A \citep{Wilson+06}, the very obvious bar and
arms in 3C\,444 \citep{Croston+11}, the `ridges' in 3C\,285 and
3C\,442A \citep{Hardcastle+07-3} and in 3C\,401 \citep{Reynolds+05}. In the best-studied cases with the clearest evidence
for ongoing strong shocks (Cygnus A and 3C\,444) the `bar' and `arm'
temperatures are consistent with the unshocked cluster temperature,
precisely as expected in our models. The agreement between our
simulations and the observations of 3C\,444, which is an FRII at $z =
0.153$ with a slightly higher jet power embedded in a slightly richer
environment, is very striking; forthcoming deep {\it XMM-Newton}
observations of this object may provide the opportunity for a detailed
test of the model's temperature predictions. This should be combined
with sensitive low-frequency observations, e.g. with LOFAR, to
establish exactly how the lobes relate to the X-ray structures.

\subsection{Suppression of cooling?}
\label{cooling}

\begin{figure*}
\includegraphics[width=14cm]{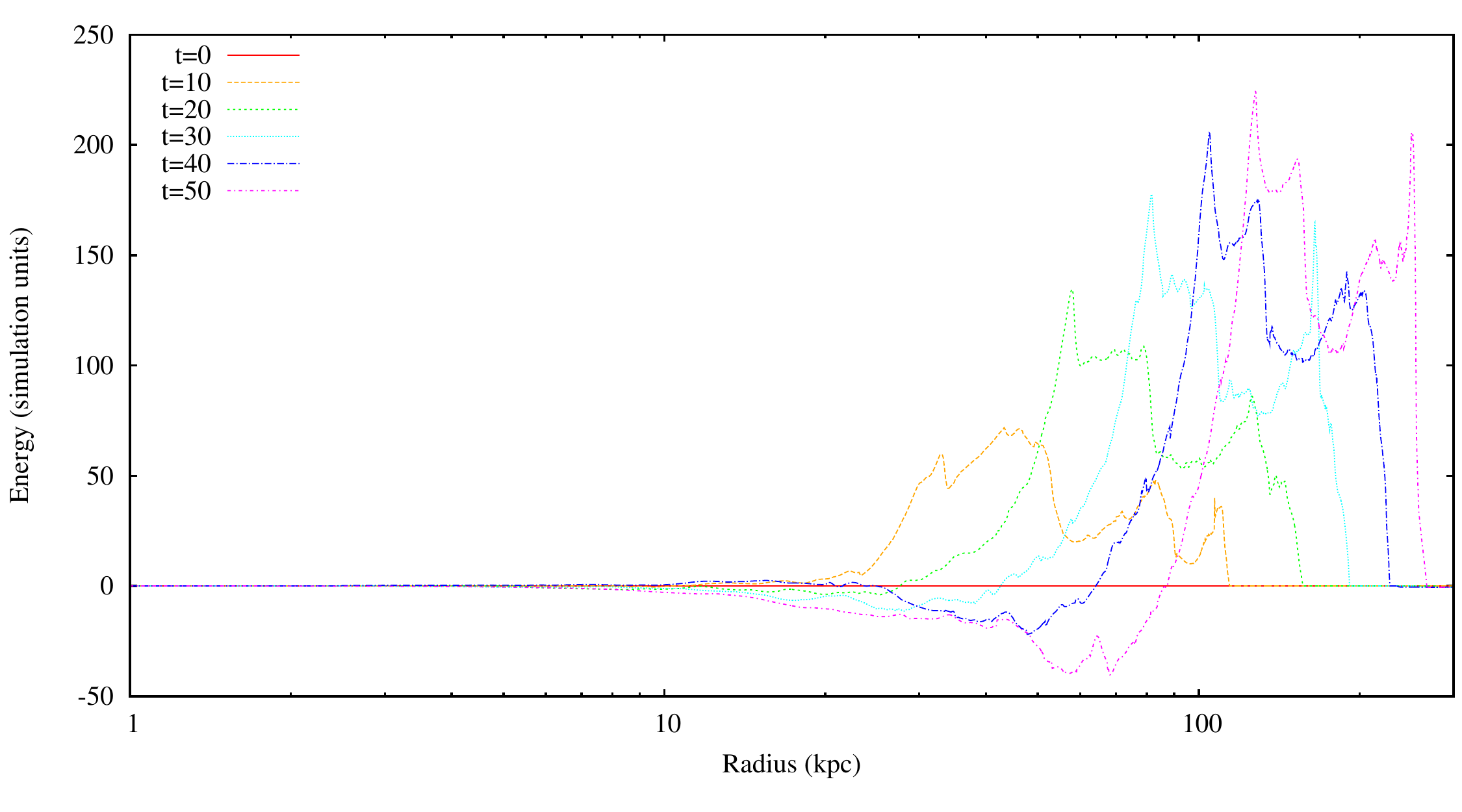}
\vskip -10pt
\includegraphics[width=14cm]{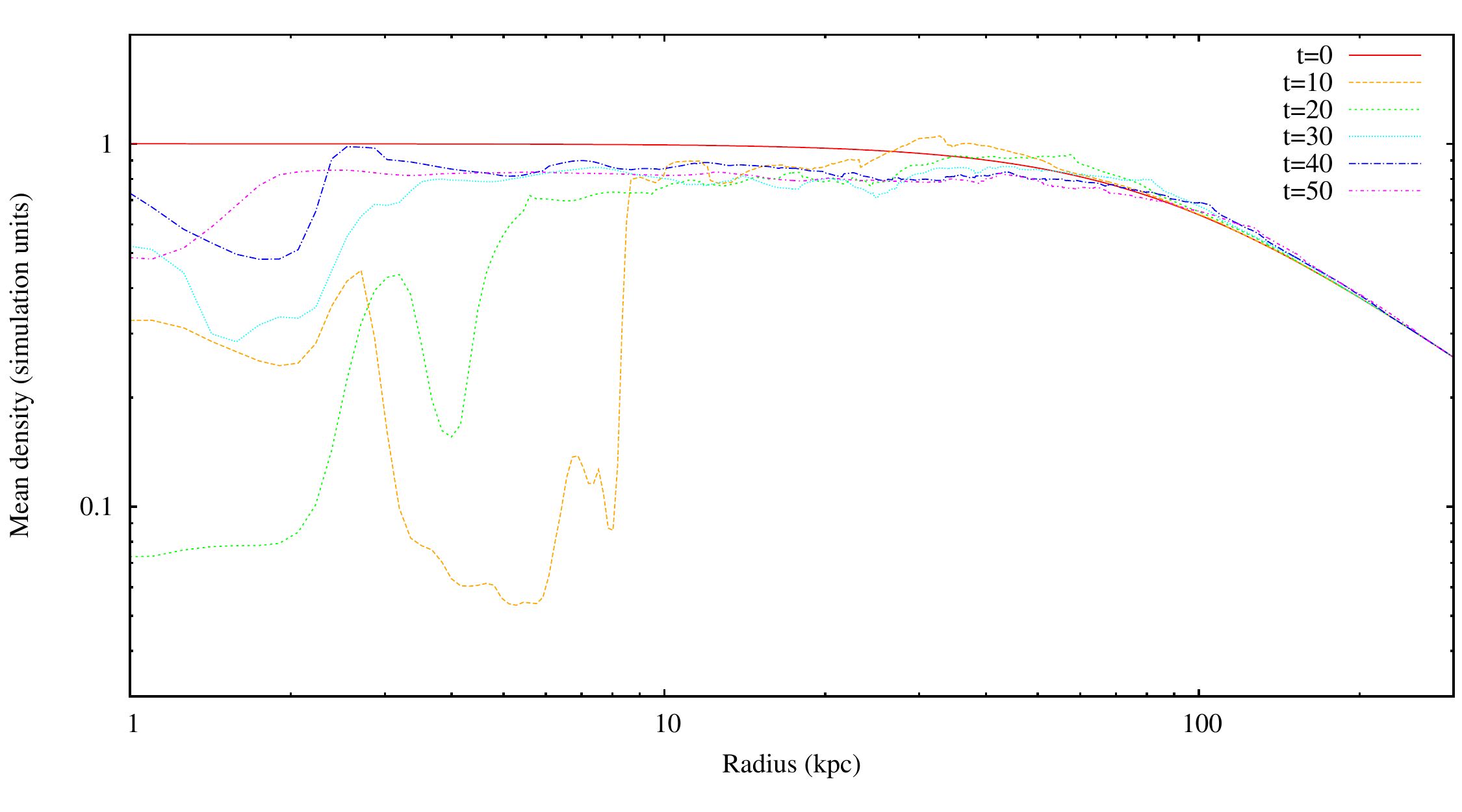}
\vskip -10pt
\includegraphics[width=14cm]{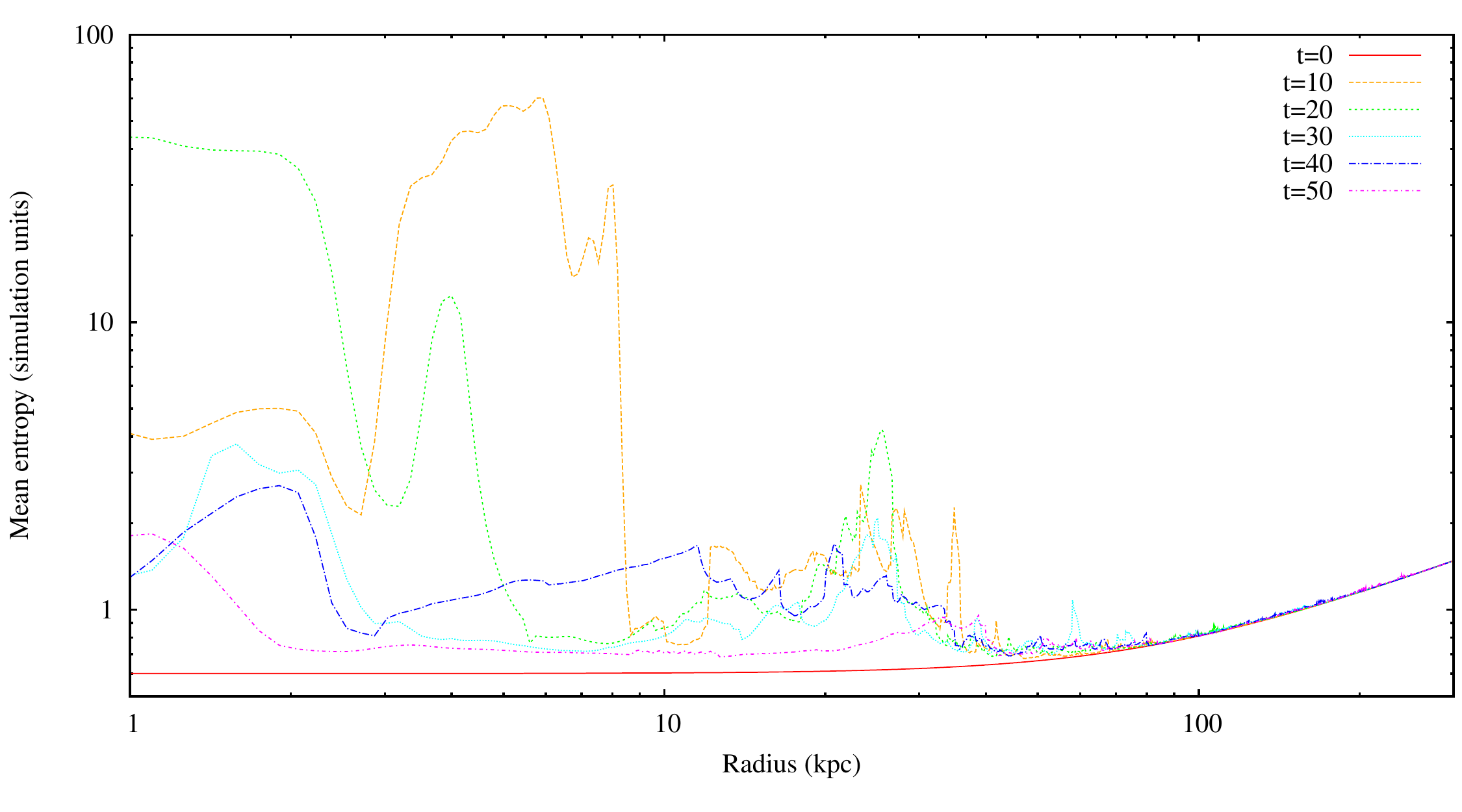}
\caption{Radial profiles of excess energy, number density and entropy
  in the external medium as a function of radius for 6 different times
  in the M25-35-40 simulation. The excess energy profile (note linear
  scale on the $y$ axis) shows total energy (thermal and kinetic) in
  all the cells that are not inside the lobe at a given radius, minus
  the value in those cells at $t=0$. The density and entropy profiles
  are the volume-weighted mean values at each radius. See Figs
  \ref{density-example} and \ref{temperature-example} for images of
  the simulated density and temperature in this simulation at these
  times ($t>0$).}
\label{profiles}
\end{figure*}

Our simulation setup was explicitly designed so that the jet power of
the FRII exceeds by roughly an order of magnitude the bolometric X-ray
luminosity of the simulated cluster environment. But this is not a
sufficient (nor indeed a necessary) condition for suppressing
radiative cooling of the X-ray-emitting material, because the cooling
rate is a strong function of density and thus of radius, and the key
requirement for prevention of the traditional cooling catastrophe is
to suppress cooling at the centre of the group/cluster. While
numerical modelling shows that powerful, FRII-like radio galaxies {\it
  can} suppress cooling \citep[e.g.][]{Basson+Alexander03}, it has been
clear for some time \citep[e.g.][]{Omma+Binney04} that they are not an {\it
  efficient} way of doing so, because the bulk of the energy/entropy
injection takes place on scales that are much larger than the cooling
radius (i.e. the radius within which the cooling time is less than
some threshold value). In sources like ours, which are intended to
come into pressure balance at the centre at late times, the situation
is even worse, because there is little or no energy input at the
centre of the cluster at all, while the increase in axial ratio with
length also means that less of the available volume of the cluster is
affected at late times. However, the cluster cores do undergo
significant heating early in the simulations.

\begin{figure*}
\includegraphics[width=16cm]{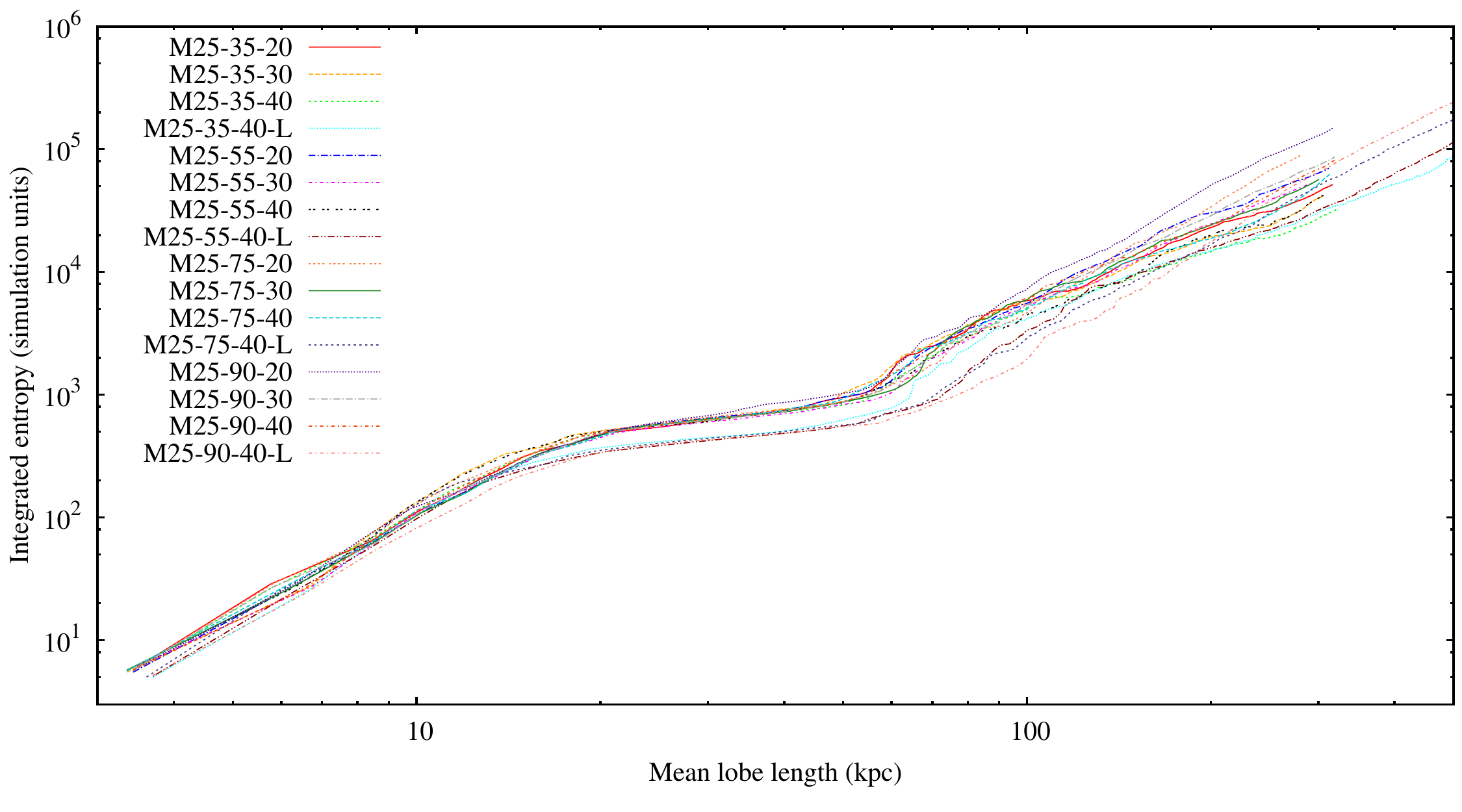}
\caption{Integrated excess entropy as a function of lobe length for the
  simulations with ${\cal M} = 25$. A baseline corresponding to
  the initial ($t=0$) entropy of the gas inside the shock front has been
  subtracted from all values plotted. The slope of this plot gives a
  measure of the efficiency with which the radio galaxy irreversibly
  heats the external medium as it grows. The three distinct regions
  here can be understood in terms of the three phases of lobe growth
  speed, and therefore shock speed (Fig.\ \ref{length}); weaker shocks
  have a lesser effect on entropy \citep{Kaiser+Binney03}.} 
\label{entropy-length}
\end{figure*}

We can quantify these effects in our simulations by looking at radial
profiles of some quantities related to cooling as a function of
simulation time. Fig.\ \ref{profiles} shows an example provided by the
M25-35-40 run, but all our results are broadly similar. What we see is
that, as expected, the total energy input to the atmosphere is mostly
co-located with the front of the lobes. The energy input to the
atmosphere at small radii is actually negative, because the effect of
the radio galaxy is to remove gas from the central regions without
providing much in the way of long-term heating in these locations.
This can be seen in the density profiles in Fig.\ \ref{profiles} -- in
the short term, the radio source has a dramatic impact on the density
of hot gas in the central regions, but as the central pressure drops
and material flows in behind the radio lobes the mean density profile
rapidly recovers to something not very dis-similar to its original
shape behind the shock front. Similarly, the entropy profiles -- here
we are plotting the entropy index, $\sigma = Pn^{-5/3}$ \citep{Kaiser+Binney03} in simulation units -- show a very marked effect at early
times but only a very modest increase in the entropy in the central
regions of the plot at late times. The effect on cooling time
(proportional to $T^{1/2}/n$: not plotted) is very similar to that on
entropy. Note also the almost completely negligible effect of the
radio source on the entropy {\it profile} at large radii -- this is
because these are volume-averaged radial profiles, and at these large
radii very little of the volume of the cluster is actually affected by
the radio galaxy.

Why do we see this behaviour of the entropy index? In our simulations,
entropy is generated at shock fronts, redistributed in mixing regions,
and conserved otherwise. The strong entropy increase in the whole core
region at early times means that this gas must in the end rise --
either buoyantly or by being carried out by the expanding radio source
-- out of the centre of the cluster; in other words, much of the core
has been efficiently heated. However, the core is then refilled by
adjacent gas from larger initial radii. Correspondingly, the density
decreases in this region. Therefore, while the effect of the jet was
to increase the entropy of much of the gas that was in the core
initially by a factor of several tens, the gas that is in the core at
the end of the simulation has only a 10-20 per cent higher entropy
than the initial conditions in this region. Because the radio galaxies
continue to drive strong shocks at large radii, they continue to
increase the entropy of the shocked gas at all times: this can be seen
in a plot of the integrated entropy, $\int \sigma {\rm d}N$, as a
function of lobe length (Fig. \ref{entropy-length}). However, as the
profiles show, this entropy increase is no longer relevant to the
cooling core, or significant over the total volume of the cluster.

We conclude that long-lived FRII radio sources with realistic dynamics
are {\it not} expected to be very efficient at suppressing cooling in
this type of environment. Though much of the energy is thermalised and
ends up in the ambient gas, only the central parts of the cluster are
efficiently, irreversibly heated, and much of this material then
leaves the central regions; at late times (i.e. $\sim 10^8$ years
after the radio activity begins), there are only very modest effects
on the central entropy and cooling time. To couple AGN output
effectively to the gas with short cooling times in the centres of
clusters, radio sources need not to grow beyond the central cluster
scale, so that they are continually in the early-time regime seen in
Fig.\ \ref{profiles}. An excellent way of doing this is to make the
radio source intermittent \citep[see e.g.][]{Mendygral+12} and
observations of the centres of cool-core clusters such as Perseus show
that this is what happens in nature most -- but not all! -- of the
time. One of the key questions to be solved in `AGN feedback' models
must be how the AGN in cool cores `know' that they are required to be
intermittent in this way while other radio galaxies, like the
classical doubles, can exhibit long-term persistent jet activity.

\section{Summary and conclusions}
\label{discussion}

We have carried out high-resolution, 2D axisymmetric hydrodynamical
simulations of the evolution of the lobes of radio galaxies in a wide
range of realistic environments. The simulations are intended to be
closely matched to the physical conditions known to obtain in the
environments of FRII radio galaxies, and to probe all size scales from
the collimation of an initially conical jet up to (and beyond) the
size scale at which the lobe pressure is expected to drop below the
pressure in the external medium. Although the details of the dynamics
internal to the lobe are probably not realistic, we argue that the
energetics of the interaction with the external medium should be
unaffected by this so long as there is a mechanism that efficiently
`thermalizes' the kinetic energy supplied by the jet, as is the case
in our simulations once the lobes grow to a few tens of simulation
lengths in size.

The key results of our work appear to be independent of both
environment and initial jet Mach number, within the range that we have
been able to probe, and can be summarized as follows:
\begin{enumerate}
\item As expected both from observations and from the design of our
  simulations, the lobes do come into pressure balance with the
  external medium in the course of the simulations (Section
  \ref{general}, Section \ref{dynamics}). At late times, the lobes are
  driven away from the centre of the host cluster by dense,
  high-pressure ICM. Thus the lobe growth is more
  consistent with Model C than with Model A of \cite{Scheuer74}, or,
  equivalently, we observe the `cocoon crushing' process of \cite{Williams91}. Lobe growth is not self-similar, and indeed axial ratios
  tend to increase with time on hundred-kpc scales, consistent with
  what is observed (Section \ref{dynamics}).
\item Unsurprisingly, given the above, the models of KA do not
  describe some of the behaviour of the simulated radio galaxies
  (Section \ref{dynamics}). The
  late-time length of the sources as a function of time is not badly
  described by the relations derived by KA, which is not surprising
  since the environment has close to a power-law behaviour of density
  with distance on these scales and the momentum flux of the jet,
  distributed over the `working surface',
  continues to drive the linear expansion of the radio source.
  However, the KA relations significantly overestimate the growth of
  lobe volume with time.
\item The pressure of the undisturbed external environment at the
  midpoint of the lobes is a reasonable, though not perfect, estimator
  of the pressure in the lobes once normal lobe growth is established
  (Section \ref{dynamics}). Lobes remain strongly overpressured with
  respect to the ambient medium at their tips. This reproduces a key
  observational result based on estimates of the lobe pressure from
  inverse-Compton observations.
\item The work done by the radio galaxy on the environment (including
  the thermal, kinetic and potential energy of the shocked gas) is
  very close to the internal energy (thermal and kinetic) of the lobes
  at all times once the lobes are established (Section
  \ref{energetics}). A simple $p_{\rm ext}V$ estimate of the work done
  by the lobes underestimates it by a factor of a few (up to an order
  of magnitude in extreme cases).
\item The simulated radio galaxies can reproduce the approximate radio
  luminosities of real 3C radio sources whose jet powers are believed
  to be comparable (Section \ref{radio}). However, there is a large
  scatter (well over an order of magnitude) in the radio powers expected from
  sources of the same power (as all our simulated sources are) in
  different environments.
\item X-ray visualization of our simulations (Section \ref{xray})
  makes some predictions for the morphological and temperature
  structure that we should observe that are in principle testable, and
  that show striking similarities to what is observed in a few
  well-studied FRII sources.
\item Finally, we note that, as has been known for a while, FRII radio
  galaxies with realistic dynamics do not couple effectively to the
  hot medium at the centres of clusters, and therefore are not in
  general effective at suppressing cooling (Section \ref{cooling}),
  though their energetic impact can be very considerable in localized
  regions around the lobes in the outer
  parts of the cluster.
\end{enumerate}

Several of these points have importance for observational work and for
models of the energetic impact of radio galaxies on their
environments. Firstly, our conclusion that the internal and external
energy of the radio galaxy remain in rough balance, independent of the
properties of the external environment or the initial jet Mach number,
means that there is a very easy recipe for estimating the work done by
an FRII radio galaxy on its environment; simply estimate the energy
stored in the lobes, ideally using an inverse-Compton observation to
measure the lobe magnetic field strength, and one also has, to within
better than a factor 2, the work done on the external environment.
Although it might be objected that our modelling does not include
important features of the physics of radio galaxies, such as magnetic
fields, which might be expected to change this conclusion, we are
reassured by the similar conclusions reached for specific simulations
with more realistic physics, including magnetic fields, by
\cite{ONeill+05} and \cite{Gaibler+09}.

Secondly, if we accept that all FRII radio galaxies follow the basic
pattern seen in our modelling and come into rough pressure balance
with their environments, independent of the initial power of their
jets, at around the lobe midpoint, once they have grown to a
sufficiently large scale (i.e. once they are in a roughly power-law
atmosphere?), then this means that large-scale FRIIs provide a probe
of the pressure of their environments; if one observes an FRII on
hundred-kpc scales, the external pressure at its midpoint should be of
order the internal pressure, which can again be assessed by
inverse-Compton observations or, with some assumptions about magnetic
field strength, from synchrotron emission alone. In the coming
decades, where deep radio observations may be far easier to get than
deep X-ray observations, FRII radio lobes may become a useful proxy for
high-redshift, high-pressure environments.

Thirdly, and on a less positive note, we draw attention to the very
large scatter seen in the radio luminosity plots. Although we have not
included radiative losses (synchrotron and inverse-Compton) in our
modelling, these are unlikely to decrease the scatter, and indeed
observational constraints on the jet power/radio power relationship
for FRIIs show similar amounts of scatter \citep[e.g.][]{Daly+12}.
While there should be a statistical correlation between jet power and
radio power, our results suggest that naive application of the results
of, e.g., \cite{Willott+99}, while they may be adequate in a statistical sense,
may give very significant errors for individual radio sources.

In future work, we hope to check the robustness of the conclusions we
have drawn here by including more of the relevant physics in the
simulations (e.g. more detailed 3D simulations, MHD, radiative
losses...) while continuing to match our modelling explicitly to
realistic radio galaxies and their environments.

\section*{Acknowledgements}

We thank Judith Croston for helpful discussions and for comments on a
draft of this paper, and an anonymous referee for helpful comments.
This work has made use of the University of Hertfordshire Science and
Technology Research Institute high-performance computing facility. The
authors acknowledge support of the STFC-funded Miracle Consortium
(part of the DiRAC facility) in providing access to the UCL Legion
High Performance Computing Facility. The authors additionally
acknowledge the support of UCL's Research Computing team with the use
of the Legion facility. MGHK acknowledges support by the cluster of
excellence `Origin and Structure of the Universe'
(http://www.universe-cluster.de/).

\bibliographystyle{mn2e}
\renewcommand{\refname}{REFERENCES}
\setlength{\bibhang}{2.0em}
\setlength\labelwidth{0.0em}
\bibliography{../bib/mjh,../bib/cards}

\end{document}